\documentclass[prd,preprintnumbers,twocolumn,eqsecnum,floatfix,letterpaper]{revtex4}
\usepackage{color}
\usepackage{calc}
\usepackage{amsmath,amssymb,graphicx}
\usepackage{amssymb,amsmath}
\usepackage{tensor}
\usepackage{bm}
\usepackage{times}
\usepackage[varg]{txfonts}
\usepackage[colorlinks, pdfborder={0 0 0}]{hyperref}
\definecolor{LinkColor}{rgb}{0.75, 0, 0}
\definecolor{CiteColor}{rgb}{0, 0.5, 0.5}
\definecolor{UrlColor}{rgb}{0, 0, 0.75}
\hypersetup{linkcolor=LinkColor}
\hypersetup{citecolor=CiteColor}
\hypersetup{urlcolor=UrlColor}
\allowdisplaybreaks
\begin{document}

\newcommand{\be}{\begin{equation}}
\newcommand{\ee}{\end{equation}}
\newcommand{\ber}{\begin{eqnarray}}
\newcommand{\eer}{\end{eqnarray}}
\def\bea{\begin{eqnarray}}
\def\eea{\end{eqnarray}}
\newcommand{\etal}{\emph{et al.}}

\newcommand{\Sl}{S_\ell}
\newcommand{\Sigmal}{\Sigma_\ell}
\newcommand{\Flux}{\mathcal{F}}
\newcommand{\LNh}{\hat{\mathbf{L}}_N}
\newcommand{\LN}{\mathbf{L}_N}
\newcommand{\bS}{\mathbf{S}}
\newcommand{\bJ}{\mathbf{J}}
\newcommand{\e}{\mathrm{e}}
\newcommand{\rmi}{\mathrm{i}}
\newcommand{\flow}{f_\mathrm{low}}
\newcommand{\fcut}{f_\mathrm{cut}}

\newcommand{\bchi}{\bm{\chi}}
\newcommand{\blambda}{\bm{\lambda}}
\newcommand{\bLambda}{\bm{\Lambda}}
\newcommand{\bchia}{\bm{\chi}_a}
\newcommand{\bchis}{\bm{\chi}_s}
\newcommand{\chis}{\chi_s}
\newcommand{\chia}{\chi_a}
\newcommand{\chiadL}{\bchia \cdot \LNh}
\newcommand{\chisdL}{\bchis \cdot \LNh}
\newcommand{\chisSqr}{\bchis^2}
\newcommand{\chiaSqr}{\bchia^2}
\newcommand{\chisDchia}{\bchis \cdot \bchia}
\newcommand{\cA}{\mathcal{A}}
\newcommand{\cB}{\mathcal{B}}
\newcommand{\cC}{\mathcal{C}}
\newcommand{\cP}{\mathcal{P}}

\newcommand{\LIGO}{LIGO Laboratory, California Institute of Technology, 
Pasadena, CA 91125, USA}
\newcommand{\CIT}{Theoretical Astrophysics, California Institute of
Technology, Pasadena, CA 91125, USA}

\title{Addressing the spin question in gravitational-wave searches: \\
Waveform templates for inspiralling compact binaries with nonprecessing spins}
\author{P.~Ajith}\email{ajith@caltech.edu}
\affiliation{\LIGO}
\affiliation{\CIT}

\begin{abstract}
This paper presents a post-Newtonian (PN) template family of gravitational waveforms from inspiralling compact binaries with non-precessing spins, where the spin effects are described by a single ``reduced-spin'' parameter.  This template family, which reparametrizes all the spin-dependent PN terms in terms of the leading-order (1.5PN) spin-orbit coupling term \emph{in an approximate way}, has very high overlaps (fitting factor $> 0.99$) with non-precessing binaries with arbitrary mass ratios and spins. We also show that this template family is ``effectual'' for the detection of a significant fraction of generic spinning binaries in the comparable-mass regime ($m_2/m_1 \lesssim 10$), providing an attractive and feasible way of searching for gravitational waves (GWs) from spinning low-mass binaries. We also show that the secular (non-oscillatory) spin-dependent effects in the phase evolution (which are taken into account by the non-precessing templates) are more important than the oscillatory effects of precession in the comparable-mass ($m_1 \simeq m_2$) regime. Hence the effectualness of non-spinning templates is particularly poor in this case, as compared to non-precessing-spin templates. For the case of binary neutron stars observable by Advanced LIGO, even moderate spins ($\LNh \cdot \bS/m^2 \simeq 0.015 - 0.1$) will cause considerable mismatches ($\sim$ 3\% -- 25\%) with non-spinning templates. This is contrary to the expectation that neutron-star spins may not be relevant for GW detection. 
\end{abstract}

\preprint{LIGO-P1100075-v4}
\maketitle

\section{Introduction}

Coalescing compact binaries consisting of stellar-mass/intermediate-mass black holes and/or neutron stars are among the most promising sources of gravitational waves (GWs) for the interferometric GW detectors like LIGO, Virgo and GEO\,600. Compact binary systems consisting of neutron stars or black holes can be produced in a variety of astrophysical scenarios, which can be broadly classified into two main classes: 1) Isolated binary evolution in which two massive stars constituting a binary undergo successive supernova explosions without disrupting the binary orbit~\cite{lrr-2006-6} 2) Dynamical formation scenarios in which two compact objects form a bound orbit due to dynamical interaction in dense stellar environments~\cite{O'Leary:2007qa}. Once formed, the binary loses orbital energy and angular momentum through GW emission and starts to inspiral and finally the binary components merge with each other. 

While spin measurements of stellar-mass black holes have indicated that many black holes may have very high spins ($||\bS||/m^2 \sim 0.2 - 0.98$)~\cite{McClintock:2011zq}, most of the observed neutron stars are found to be weakly spinning~\cite{2005AJ....129.1993M}. Additionally, for neutron stars, there is a theoretical upper limit ($||\bS||/m^2 \sim 0.7$) on the spin rate beyond which the neutron star is gravitationally unstable~\cite{1994ApJ...424..823C,Lattimer:2006xb}. In the case of compact binaries formed in isolated evolution, it is likely that the spins will be nearly aligned to the orbital angular momentum~\cite{Kalogera:2004pr}. But the distribution of the spin tilt angles (angle between the spin and orbital angular momentum) depends on the distribution of supernova kicks~\cite{1996Natur.381..584K}.  On the other hand, in the case of binaries formed in dynamical interactions, there is no prior reason to expect any spin alignments. If the spins are misaligned with the orbital angular momentum, the general relativistic spin-orbit and spin-spin coupling will cause the spins to precess around the (nearly fixed) direction of the total angular momentum~\cite{PhysRevD.49.6274}. The complexity of the binary's dynamics will be encoded in the GW signals observed by a detector. 

The GW signals from compact binaries buried in the noisy data of interferometric detectors are best extracted by employing the technique of \emph{matched filtering}, which involves cross correlating the data with theoretically calculated templates of the expected GW signals.    
Theoretical GW templates can be constructed by solving Einstein's equations using analytical approximation methods and/or numerical techniques. When the compact objects are well separated and are slowly moving ($v/c \ll 1$) gravitational waveforms can be computed using the post-Newtonian (PN) approximation to General Relativity~\cite{Blanchet:LivRev}. The PN approximation breaks down as the compact objects reach the ultra-relativistic regime, and an accurate description of the merger process requires exact solutions of Einstein's equations which can only be obtained by large-scale numerical simulations~\cite{Pretorius:2005gq,Campanelli:2005dd,Baker05a}.

Analytical GW templates (parameterized by the masses and spins) can be constructed by combining PN calculations with numerical-relativity simulations~(see, e.g., \cite{Buonanno:2007pf,Damour:2009kr,Ajith:2007qp,Ajith:2007kx,Santamaria:2010yb,Sturani:2010ju}). In the case of ``low-mass'' ($m_1 + m_2 \lesssim 12 M_\odot$~\cite{Ajith:2007xh,Robinson:2008un,Buonanno:2009zt}) binaries where the signal-to-noise ratio (SNR) is almost entirely contributed by the inspiral portion of the signal, it is sufficient to model only the inspiral accurately, where the PN approximation holds. 
Although the expected form of the signals can be computed as a function of the source parameters, the parameters of a signal that is buried in the noise is not known \emph{a priori}. Thus the data is cross correlated with a ``bank'' of theoretical templates corresponding to different (astrophysically plausible) values of physical parameters. A geometrical formalism has been developed for ``laying down'' templates in the parameter space of compact binaries~\cite{Sathyaprakash:1991mt,Owen:1995tm}. Parameters of the templates are chosen in such a way that the loss of SNR due the mismatch between two neighboring templates is less than an acceptable value, while at the same time keeping the total number of templates employed in the search computationally tractable.  

However, templates for spinning binaries are characterized by a large number of parameters (two for the masses, six for the spins, two for the inclination and polarization angles), complicating placement of templates in the bank and considerably increasing the computational cost of searches. Furthermore, many different spin configurations are known to be degenerate, making it unnecessary to employ templates corresponding to \emph{all} parameters. 

The idea of constructing an \emph{effective} template family parameterized by a smaller number of parameters that has high enough overlaps with the expected signals was first proposed by Apostolatos~\cite{PhysRevD.54.2438}, who introduced a modulational sinusoidal term in the frequency-domain phase of the templates to capture the oscillatory effects of precession. However, it was found that this template family fails to capture the target signals with sufficient efficiency~\cite{PhysRevD.67.042003}. Later, Buonanno, Chen and Vallisneri (BCV)~\cite{BCV2} proposed a phenomenological template family that has better overlaps with the target signals, by introducing modulation effects in both the frequency-domain- amplitude and phase of the templates. Several of these phenomenologically introduced parameters could be searched over analytically, thus reducing computational costs significantly. A search for GWs from spinning binaries was performed using the data from LIGO's third science run, employing BCV templates~\cite{PhysRevD.78.042002,Jones:2010iv}. Interestingly, the sensitivity of this search towards spinning binaries was found to be not any better than a search employing non-spinning templates. Indeed, BCV have cautioned that the increased degrees of freedom associated with the detection statistic (due to the analytical maximization of several phenomenological parameters) will also increase the false alarm rate. Later Van Den Broeck~\etal~\cite{VanDenBroeck:2009gd} demonstrated that, while BCV templates have high overlaps with the target signals, the increased degrees of freedom associated with the detection statistic and the lack of good methods for vetoing non-Gaussian detector glitches that mimic the expected signals (``signal-based vetoes'') resulted in producing a much higher false alarm rate in the presence of non-Gaussian noise, and hence negated the advantages. They argued that the standard non-spinning frequency-domain 3.5PN templates have higher detection efficiency (at a given false alarm rate) towards spinning binaries than the current implementation of the phenomenological BCV template bank. 

BCV~\cite{BCV2} also proposed a physical template family, which presumably will not suffer from this limitation. This template family, which assumes that only one compact object has significant spin, is parametrized by four parameters that need to be searched over (two masses, magnitude of the single spin, spin tilt angle). This template family was extensively studied by Pan~\etal~\cite{Pan:2003qt} and Buonanno~\etal~\cite{Buonanno:2004yd}, where they demonstrated that the template family is \emph{effectual}~\cite{DIS98} in detecting single-spin as well as double-spin binaries using Initial detectors. They also demonstrated the possibility of reducing the number of spin parameters to one in certain regions in the parameter space. This template family has been implemented in the LIGO-Virgo search pipelines~\cite{FaziThesis,Harry:2010fr,Harry:2011qh}. Another enhancement of the non-spinning frequency-domain template family which extends the symmetric mass ratio ($\eta = m_1 m_2/m^2$) to unphysical values ($\eta > 0.25$) has also been proposed as a detection template family for spinning binaries~\cite{Boyle:2009dg,Aylott:2009ya,Lundgren}. 

In this paper we propose a frequency-domain PN template family characterized by a \emph{single} ``reduced-spin'' parameter (and two masses) for the case of binaries with non-precessing spins (spins aligned/anti-aligned with the orbital angular momentum). This template family, which reparametrizes all the spin-dependent PN terms in terms of the leading-order (1.5PN) spin-orbit coupling term \emph{in an approximate way}, has very high overlaps with non-precessing binaries with arbitrary mass ratios and spins. We also show that this ``reduced-spin'' template family is able to capture a significant fraction of generic precessing binaries in the comparable-mass regime ($q \equiv m_2/m_1 \lesssim 10$). This might provide an efficient and feasible way of searching for spinning low-mass binaries in the comparable-mass regime.  This result also shows that the secular (non-oscillatory) spin-dependent effects in the phase evolution (which are taken into account by the non-precessing templates) are more important than the oscillatory effects in the case of (nearly) equal-mass binaries even with generic spins. 

This paper is organized as follows: Section~\ref{sec:summary} summarizes the main findings of this paper as well as lists the limitations of this work. Section~\ref{sec:targetWaves} provides a description of the PN waveforms from precessing binaries computed in the adiabatic approximation, which are assumed to be the ``target signals'' that we want to detect. Section~\ref{sec:Detection} is a brief introduction to the GW data analysis of inspiralling compact binaries. In Section~\ref{sec:EffSpinTempl}, we construct the frequency-domain PN template family parametrized by a reduced spin parameter and demonstrate the \emph{effectualness}~\cite{DIS98} of the template family in detecting binaries with non-precessing spins. In Section~\ref{sec:Precession} we investigate the effectualness of the template family in detecting precessing binaries. We use geometrical units throughout the rest of the paper: $G = c = 1$.  

\section{Summary of results, recommendations, and limitations of this work}
\label{sec:summary}

A brief summary of the main findings of this paper is given below: 

\begin{enumerate}

\item For the case of binary neutron stars observable by Advanced LIGO, even moderate spins ($\LNh \cdot \bS/m^2 \simeq 0.015 - 0.1$) will cause significant mismatches (3\% -- 25\%) with non-spinning templates. This is contrary to the expectation that spin effects may not be relevant for the detection of GWs from binary neutron stars. Secular spin-dependent effects in the phase evolution are more important than oscillatory effects in the case of comparable-mass ($m_1 \simeq m_2$) binaries. Hence the effectualness of non-spinning templates is particularly poor in the comparable-mass regime, as compared to templates describing non-precessing spins, which take into account the secular effects (see Figs.~\ref{fig:FFHists} and \ref{fig:PrecFFScatterPlotsNS}). 

\item In the case of binaries with \emph{non-precessing} spins, it is possible to describe the dominant spin effects in terms of a single ``reduced-spin'' parameter (the combination of spins and mass ratio that describes the leading-order spin-orbit-coupling term $\beta$~\cite{Poisson:1995ef}). The reduced-spin template family has very high fitting factor~\cite{Apostolatos:1995pj} towards non-precessing binaries with arbitrary spins and mass ratios (see Fig.~\ref{fig:TF2TF2ReduceFF}), while the non-spinning template family can cause significant loss of SNR in certain regions of parameter space, e.g., spins aligned to the angular momentum (see Fig.~\ref{fig:TF2TF2ReduceMatch}). Note that binaries with spins nearly aligned to the angular momentum are expected from isolated binary evolution~\cite{Kalogera:2004pr}.

\item Assuming a target population of binaries with uniformly distributed spin magnitudes ($||\bS||/m^2$) in the interval (0, 0.98) for black holes and (0, 0.7) for neutron stars, and isotropically distributed spin angles and inclination-polarization angles, we show that almost the entire population of \emph{equal-mass} binaries with \emph{generic spins} can be detected using the reduced-spin templates with very high fitting factors. On the other hand, only $\sim 51\% \, (52\%) \, 57\%$ of the equal-mass population with total mass $2.8 M_\odot \,(12 M_\odot)\, 20 M_\odot$ produces the same fitting factor with non-spinning PN templates. Even if we restrict the neutron star spins to $(0,0.3)$, this result does not change significantly (see Fig.~\ref{fig:FFHists}). 

\item The effectualness of the reduced-spin templates decreases with increasing mass ratio ($m_2/m_1 \gg 1$), since for binaries with significantly unequal masses, oscillatory effects of precession become crucial. Still, the reduced-spin template family performs considerably better than non-spinning templates: $\sim$ 60\% (45\%) 37\% of the population with $m_2/m_1 = 3.5 \, (7.1) \, 13.3$ produces fitting factor $>$ 0.97 with reduced spin templates, while $\sim$ 44\% (36\%) 32\% produces the same fitting factor with non-spinning templates (see Fig.~\ref{fig:FFHists}). 

\item In the case of unequal-mass binaries, most of the binaries producing fitting factor $<0.9$ with the reduced spin templates are significantly tilted with respect to the detector and have spins of the more massive object highly non-aligned with the orbital angular momentum. In the case of GW searches where astrophysical priors restrict the inclination angles or spin tilt angles of the target population to small values (e.g., binary mergers producing short-hard gamma ray bursts observable to us), it might suffice to consider only the secular spin effects (see Fig~\ref{fig:PrecFFScatterPlotsEffSpin}). 

\item For highly unequal-mass binaries, spin effects are almost entirely determined by the spin of the more massive compact object~\footnote{Note that for a black hole with Kerr parameter $\chi$, the spin angular momentum scales with the square of the mass.}, and hence a template family describing precessional effects assuming only one spinning compact object, such as the physical template family proposed by BCV~\cite{BCV2,Pan:2003qt,FaziThesis,Harry:2010fr} should be able to model these binaries accurately. This suggests one natural way of splitting the parameter space in GW searches: Binaries in the comparable-mass regime (where precession effects are not significant) can be detected employing templates described by a single spin parameter, while the detection of binaries in the high mass ratio regime (where precession effects are significant) will require templates described by two spin parameters.   

\item The observation that binaries in the comparable-mass regime can be detected employing a template family described by a single spin parameter also suggests that it will be hard to estimate the other spin parameters of the binary accurately employing restricted PN templates. This underlines the need of including all the known physical information (from higher harmonics, merger-ringdown etc.) in the parameter-estimation pipelines which might help to disentangle the correlations between different spin components. 

\end{enumerate}

A note on the limitations of this work: We consider only the dominant harmonic of the gravitational waveforms from the inspiral stage of the binary coalescence, and the amplitude of the waveforms is computed only in the leading order (``restricted'' PN approximation). In the detection problem, higher harmonics are believed to be not important for the case of low-mass binaries in the comparable mass regime. But for more massive and highly unequal-mass binaries, the contribution from higher harmonics can be significant -- especially if the binary is tilted with respect to the detector~\cite{Arun:2009,VanDenBroeck:2006ar}. Also, the effect of post-inspiral stages is negligible in the case of low-mass binaries ($m \lesssim 12 M_\odot$), but can not be neglected in the case of high-mass binaries. Still, we consider binaries with $m$ as large as $20 M_\odot$ in this paper. The rationale for this choice is that in this paper we are investigating only the spin-effects in the GW detection problem. Since we are neglecting the effects of post-inspiral stages, we evaluate the performance of different template families only in the frequency range $f \leq f_\mathrm{ISCO}$, where $f_\mathrm{ISCO}$ is the frequency of the innermost stable circular orbit (ISCO) in the Schwarzschild geometry. Note that, for the case of spinning binaries, the ``actual'' ISCO frequency depends on the spin; we ignore this. 

Additionally, in this paper, we quantify the effectualness of the template family only in terms of the fitting factor, which quantifies the loss of SNR due to the mismatch of the signal and template in Gaussian noise. Since actual detector data is not perfectly Gaussian, additional signal-based vetoes are used in the detection statistic, such as the ``chi-square veto''~\cite{Allen:2004gu}, whose effects are not considered here. But, note that, unlike the case of the BCV phenomenological template family, signal-based vetoes can be readily implemented for the case of the reduced-spin template family. But we note that a complete characterization of the effectualness of a template family requires detailed studies using actual detector data. This is beyond the scope of this paper. 

\smallskip 

\section{Post-Newtonian waveforms in the adiabatic approximation}
\label{sec:targetWaves}

In this discussion, we will only consider binaries in quasi-circular~\footnote{Note that the orientation of the orbital plane can change during the evolution as a result of the spin precession. Thus these orbits are best referred to as ``quasi-spherical'' orbits.} orbits. Indeed, inspiralling compact binaries are expected to lose their eccentricity before their GW signals reach the sensitive frequency band of ground-based detectors~\cite{Peters:1964zz,Peters:1963ux}. In the early stages of the evolution of the binary due to GW emission, the change in orbital frequency is much smaller than the orbital frequency itself. During this \emph{adiabatic} inspiral, the loss of the specific orbital binding energy $E(v)$ (binding energy per unit mass) is related to the energy flux of gravitational radiation $\Flux(v)$ in the following way: $m\, dE(v)/dt = -\Flux(v)$. 
This ``energy balance'' argument provides the following coupled ordinary differential equations from which orbital phase evolution $\varphi(t)$ of the binary can be computed 
\begin{eqnarray}
\label{eq:phasingformula}
\frac{d \varphi}{dt} = \frac{v^3}{m}\,, ~~~~~~~ \frac{dv}{dt} = - \frac{\mathcal{F}(v)}{m\,E'(v)}\,.
\end{eqnarray}
Above, $v$ is a velocity parameter which is related to the orbital frequency $\omega$ by $v \equiv (m \omega)^{1/3}$ where $m \equiv m_1+m_2$ is the total mass of the binary, and $E'(v) \equiv dE(v)/dv$. The energy and flux functions can be computed as PN expansions in terms of the small parameter $v$. Currently the energy function has been computed up to 3PN order $(v^6)$~~\cite{Arun:2009, Jaranowski:1999ye, Poisson:1997ha, Kidder:1995zr, Blanchet:2002, Blanchet:2005a, Blanchet:2005b, Blanchet:2006gy, Blanchet:2007, Blanchet:2010, Faye:2006gx, WillWiseman:1996, Blanchet:2000ub, deAndrade:2000gf,Damour:2001bu} and flux function up to 3.5PN order $(v^7)$~~\cite{Arun:2009, Faye:2006gx, Blanchet:2007, Mikoczi:2005dn, Racine:2008kj, Blanchet:2004ek, Blanchet:2006gy}; but the spin effects have been computed only up to 2PN and 2.5PN orders, respectively~\footnote{While this paper was being prepared Blanchet, Buonanno and Faye~\cite{Blanchet:2011zv} have also computed the spin-orbit effects appearing at 3PN order in the flux function. This paper does not include those terms.}. 

The energy balance equation can be solved in a number of different ways (see~\cite{Buonanno:2009zt} for a recent overview) which are perturbatively equivalent at the corresponding PN order. But, since the PN expansion of the energy and flux functions is known up to only a limited PN order, the actual results of these different methods can be somewhat different. In this paper, we construct a new approximant to the GW phasing computed by re-expanding the rational function $E'(v)/\Flux(v)$ and truncating it at 3.5PN order (spin terms are considered only up to 2.5PN):
\begin{widetext}
\begin{align}\label{eq:EbyF}
\frac{E'(v)}{\Flux(v)} & = \frac{-5}{32 v^9 \eta} \left\{1 + v^2 \left[\frac{11 \eta }{4}+\frac{743}{336} \right] 
    + v^3 \left[\frac{113}{12} \left(\biggl( 1-\frac{76\eta}{113}\biggr) \, \chisdL + \delta \, \chiadL \right)-4 \pi \right] \right. \nonumber \\ 
    &\quad + v^4 \left[(\chiadL)^2 \left(30 \eta -\frac{719}{96}\right)-\frac{719 \chiadL \, \chisdL \delta
   }{48}+\chiaSqr \left(\frac{233}{96}-10 \eta \right)+\frac{233 \chisDchia \delta}{48} \right. \nonumber\\
   &\qquad \left. +(\chisdL)^2 \left(-\frac{\eta }{24}-\frac{719}{96}\right)+\chisSqr \left(\frac{7 \eta
   }{24}+\frac{233}{96}\right)+\frac{617 \eta ^2}{144}+\frac{5429 \eta }{1008}+\frac{3058673}{1016064} \right] \nonumber \\ 
    &\quad + v^5 \left[\chiadL \delta  \left(\frac{7 \eta }{2}+\frac{146597}{2016}\right)+\chisdL \left(-\frac{17 \eta
   ^2}{2}-\frac{1213 \eta }{18}+\frac{146597}{2016}\right)+\frac{13 \pi  \eta }{8}-\frac{7729 \pi }{672} \right]  \nonumber\\ 
    &\quad + v^6 \left[\frac{1712 \gamma_E }{105}+\frac{25565 \eta ^3}{5184}-\frac{15211 \eta
   ^2}{6912}-\frac{451 \pi ^2 \eta }{48}+\frac{3147553127 \eta }{12192768}+\frac{32 \pi
   ^2}{3}-\frac{10817850546611}{93884313600}+\frac{1712 \ln (4v)}{105} \right]  \nonumber\\ 
    &\quad \left. + v^7 \left[\frac{14809 \pi  \eta ^2}{3024}-\frac{75703 \pi  \eta }{6048}-\frac{15419335 \pi }{1016064} \right] \right\}. 
\end{align}
\end{widetext}
Above, $\eta \equiv m_1 m_2/m^2$ is the symmetric mass ratio, $\delta \equiv (m_1-m_2)/m$ the asymmetric mass ratio, $\gamma_E$ the Euler's constant, and $\LNh$ the unit vector along the Newtonian orbital angular momentum $\LN \equiv m^2\eta/v ~ \LNh$. The spin variables $\bchis$ and $\bchia$ are related to the (dimensionless) spin vectors of the binary as: 
\begin{equation}
\bchis = (\bchi_1 + \bchi_2)/2, ~~~ \bchia = (\bchi_1 - \bchi_2)/2,
\end{equation}
where $\bchi_i \equiv \bS_i/m_i^2$, $\bS_i$ being the spin angular momentum of the $i$th compact object. 

Equation~(\ref{eq:EbyF}) can be plugged back to Eq.~(\ref{eq:phasingformula}) to get an explicit expression of the orbital phase in terms of $v$: 
\begin{widetext}
\begin{align}\label{eq:phiOfV}
\varphi(v) & = \varphi_0 - \frac{1}{32 v^5 \eta} \left\{1 + v^2 \left[ \frac{55 \eta }{12}+\frac{3715}{1008} \right] 
        + v^3 \left[\frac{565}{24} \left(\biggl( 1-\frac{76\eta}{113}\biggr) \, \chisdL + \delta \, \chiadL \right) -10 \pi \right] \right. \nonumber \\ 
    &\quad + v^4 \left[(\chiadL)^2 \left(150 \eta -\frac{3595}{96}\right)-\frac{3595 \chiadL \, \chisdL \delta
   }{48}+\chiaSqr \left(\frac{1165}{96}-50 \eta \right)+\frac{1165 \chisDchia \delta }{48} \right.  \nonumber\\ 
   &\qquad \left. +(\chisdL)^2 \left(-\frac{5 \eta }{24}-\frac{3595}{96}\right)+\chisSqr \left(\frac{35 \eta
   }{24}+\frac{1165}{96}\right)+\frac{3085 \eta ^2}{144}+\frac{27145 \eta }{1008}+\frac{15293365}{1016064} \right]  \nonumber\\ 
    &\quad + v^5 \left[\left(\chiadL \left(-\frac{35 \delta  \eta }{2}-\frac{732985
   \delta }{2016}\right)+\chisdL \left(\frac{85 \eta ^2}{2}+\frac{6065 \eta
   }{18}-\frac{732985}{2016}\right)-\frac{65 \pi  \eta }{8}+\frac{38645 \pi }{672}\right) \ln (v) \right]  \nonumber\\ 
    &\quad + v^6 \left[-\frac{127825 \eta ^3}{5184}+\frac{76055 \eta ^2}{6912}+\frac{2255 \pi ^2 \eta }{48}-\frac{15737765635 \eta
   }{12192768}-\frac{1712 \gamma_E }{21}-\frac{160 \pi
   ^2}{3}+\frac{12348611926451}{18776862720}-\frac{1712 \ln (4v)}{21} \right]  \nonumber\\ 
    &\quad \left. + v^7 \left[-\frac{74045 \pi  \eta ^2}{6048}+\frac{378515 \pi  \eta }{12096}+\frac{77096675 \pi }{2032128} \right] \right\}, 
\end{align}
\end{widetext}
where $\varphi_0$ is a certain reference phase. 

Thus the phasing formula can be reduced to one differential equation describing the evolution of the orbital frequency, and one explicit expression of the orbital phase: 
\begin{eqnarray}
\label{eq:phasingformula2}
\frac{dv}{dt} = \left[\frac{-mE'(v)}{\Flux(v)}\right]^{-1}\,,~~~ \varphi(v) & = \varphi_0 - {\displaystyle \frac{1}{32 v^5 \eta} \, \Biggl\{1 + \dots\Biggr\}}. 
\end{eqnarray}
We call this particular way of solving the phasing formula the ``TaylorT5'' approximant~\footnote{Note that this particular way of writing $\varphi(v)$ as an explicit expansion in terms of $v$ is widely used in the PN literature (see e.g.,~\cite{Blanchet:2006gy}).}. The choice of this particular approximant as the target signal family is motivated by the following reason: Since the spin-dependent terms in the PN expansion of the energy and flux functions are available only up to a rather low 2.5PN order, the different approximants give somewhat different results. We want to isolate this issue from the issue of the effect of spin-precession in the target signals. Thus, we construct the target waveforms in such a way that they are as close to the non-precessing frequency-domain template family as possible \emph{in the limit of non-precessing spins}. Since the frequency-domain template family used in this paper (``TaylorF2'' approximant~\cite{Buonanno:2009zt}) is constructed based on a re-expansion of $E'(v)/\Flux(v)$, we choose to construct the time-domain target waveforms also based on this re-expansion. Note that the effectualness of the template family (although weakly) depends on the particular approximation used in the construction of the target and template waveforms. This is an indication of the level of truncation error in the PN expansion, and points to the need of computing the higher PN order spin  terms. This is being explored in an ongoing work~\cite{Ajith:2011xx}. Also note that the ``TaylorT5'' approximant has an additional advantage (over TaylorT1 and TaylorT4) that only one differential equation needs to be solved numerically; the orbital phase is computed as an explicit expansion in $v$. 

If the spin vectors are misaligned with the orbital angular momentum, the spin-orbit and spin-spin coupling cause the spins and orbital angular momentum to precess around the nearly constant direction of the total angular momentum $\bJ$, constantly changing the angle between the spins and angular momentum~\cite{PhysRevD.49.6274}. The evolution equations for the orbital angular momentum and spins, including the next-to-leading-order spin-orbit terms, are given by~\cite{BCV2,Blanchet:2006gy}: 
\begin{equation}
\frac{m^2\eta}{v}\left[1 + \left(\frac{3}{2} + \frac{\eta}{6}\right)v^2 \right] \, \frac{d \LNh}{dt}  =  -\frac{d}{dt} (\bS_1 + \bS_2),
\label{eq:precessionEqnsL}
\end{equation}
and 
\begin{equation}
\frac{d \bS_i}{dt}  =  \bm{\Omega}_i \times \bS_i \,,~ i = 1,2, 
\label{eq:precessionEqnsS}
\end{equation}
where
\begin{align}
\label{eq:Omega}
\bm{\Omega}_1 & =  \frac{v^5}{m} \left\{\left(\frac{3}{4}+\frac{\eta}{2}-\frac{3\delta}{4}\right) \LNh \right. \nonumber \\
     & \quad +  \frac{v}{2m^2} \left[-3 \, (\bS_2 + q \, \bS_1).\LNh ~ \LNh + \bS_2 \right]  \nonumber \\
     & \quad +  \left. v^2 \left(\frac{9}{16} + \frac{5\eta}{4} - \frac{\eta^2}{24} - \frac{9\delta}{16} + \frac{5\delta \eta}{8} \right)  \LNh  \right\}, \nonumber \\
\bm{\Omega}_2 & =  \frac{v^5}{m} \left\{\left(\frac{3}{4}+\frac{\eta}{2}+\frac{3\delta}{4}\right) \LNh \right. \nonumber \\
     & \quad +  \frac{v}{2m^2} \left[-3 \, (\bS_1 + q^{-1} \, \bS_2).\LNh ~ \LNh + \bS_1 \right]  \nonumber \\
     & \quad +  \left. v^2 \left(\frac{9}{16} + \frac{5\eta}{4} - \frac{\eta^2}{24} + \frac{9\delta}{16} - \frac{5\delta \eta}{8} \right)  \LNh  \right\}. 
\end{align}
Above, $q \equiv m_2/m_1$ is the mass ratio. The instantaneous precession frequency of the individual spins is $||\Omega_i||$.

The orbital frequency, spins and orbital angular momentum can be evolved by solving the differential equations Eqs.~(\ref{eq:phasingformula2}), (\ref{eq:precessionEqnsL}) and (\ref{eq:precessionEqnsS}). In order to perform the evolution, we adopt the coordinate system proposed by Finn and Chernoff~\cite{PhysRevD.47.2198}, and subsequently used by several authors. The $z$-axis of this (fixed) source coordinate system is determined by the initial total angular momentum vector $\bJ \simeq \LN + \bS_1 + \bS_2$ and the $x$-axis is chosen in such a way that the detector lies in the $x-z$ plane. 

\begin{figure}[tb]
\begin{center}
\includegraphics[width=3.0in]{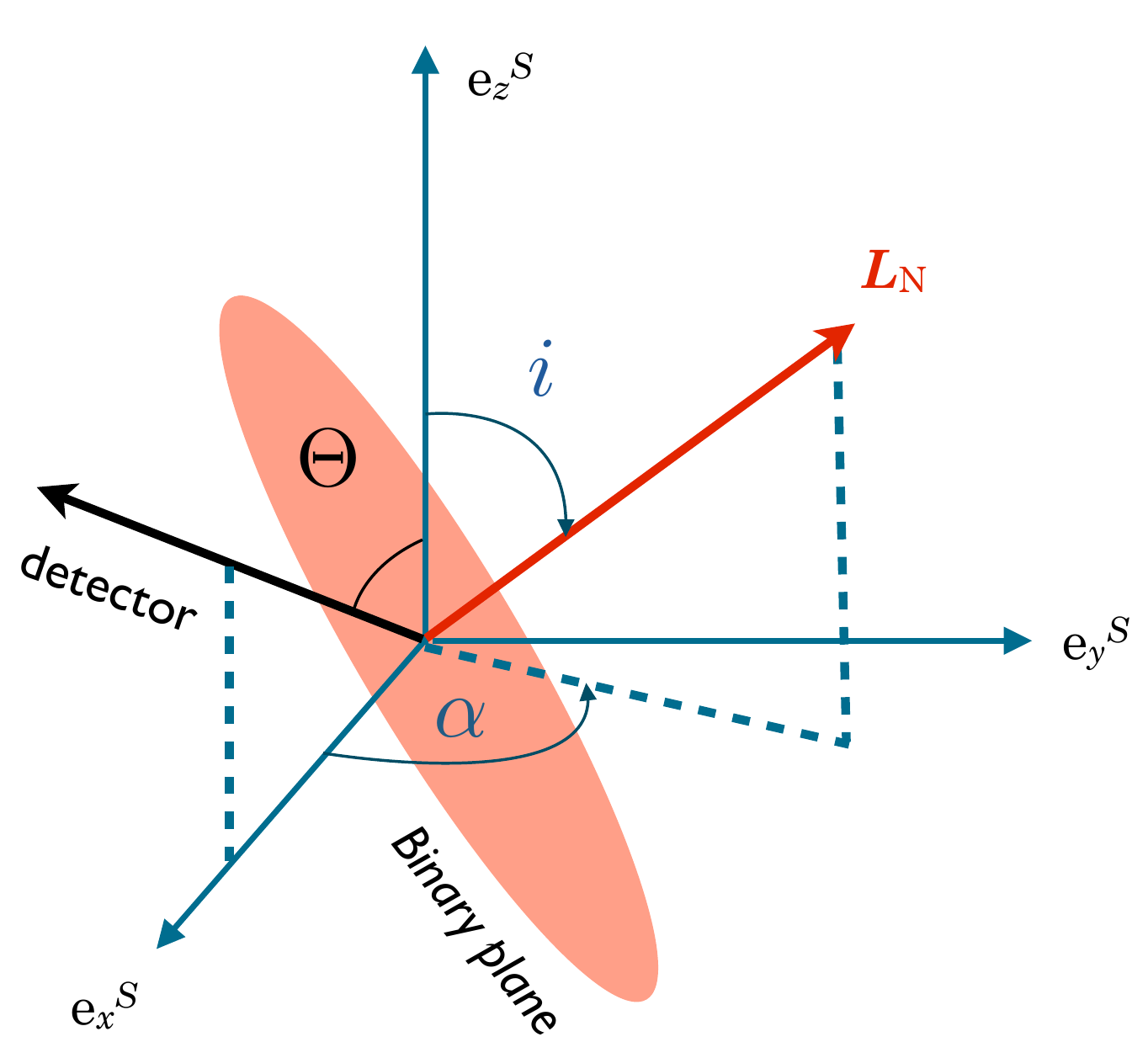}
\caption{Source frame in the Finn-Chernoff convention. The $z$-axis points to the direction of the initial total angular momentum vector $\bJ$, the detector lies in the $x-z$ plane, and the angles $\alpha$ and $i$ describe the evolution of the orbital angular momentum vector $\LN$ in the source frame.}
\label{fig:SourceFrame}
\end{center}
\end{figure}

In the absence of precession, the phase evolution of the (dominant harmonic) GWs is twice the orbital phase $\varphi(t)$. But the precession of the orbital plane introduces additional modulation in the orbital phase. If we define $\Phi(t)$ as the orbital phase with respect to the line of ascending nodes (the point at which the orbit crosses the $x-y$ plane from below) then the the phase evolution of the (dominant harmonic) GWs is given by $2\Phi(t)$, where $\Phi(t)$ is given by 
\begin{equation}
\label{eq:dPhiByDt}
\frac{d\Phi}{dt} = \omega - \frac{d\alpha}{dt} \cos i, 
\end{equation}
where $\alpha$ and $i$ are the angles describing the evolution of the orbital angular momentum vector $\LNh$ in the Finn-Chernoff coordinate system:
\begin{equation}
\alpha \equiv \arctan (\hat{L}_{Ny}/\hat{L}_{Nx}), ~~~ i \equiv \arccos (\hat{L}_{Nz}). 
\end{equation}
Computation of gravitational waveforms in the detector frame follows the description of BCV~\cite{BCV2} (Section IIC). In the \emph{restricted} PN approximation, the resulting gravitational waveform observed at the detector can be written as: 
\begin{equation}
\label{eq:cqsqwaveforms}
h(t) = C_Q(t)\,\cos 2\left[\Phi(t)+\Phi_0\right] + S_Q(t)\,\sin 2\left[\Phi(t)+\Phi_0\right]\,,
\end{equation}
where
\begin{eqnarray}
\label{eq:cqeq}
C_Q(t) &=& -\frac{4 \eta m}{D}\,v^2\,\left[ C_+(t)\,F_+ + C_\times(t)\,F_\times \right ]\,, \nonumber \\
S_Q(t) &=& -\frac{4 \eta m}{D}\,v^2\,\left[ S_+(t)\,F_+ + S_\times(t)\,F_\times \right ]\,,
\end{eqnarray}
and $\Phi_0$ is a constant that depends on the initial configuration of the binary. Above, $D$ is the luminosity distance to the binary, while $F_+$ and $F_\times$ are the antenna pattern functions: 
\begin{eqnarray}
F_+ &=& \frac{1}{2}(1+\cos^2 \theta)\,\cos 2\phi\,\cos 2 \psi -
\cos \theta\,\sin 2\phi\,\sin2\psi \,, \nonumber \\
F_\times &=& \frac{1}{2}(1+\cos^2 \theta)\,\cos 2\phi\,\sin 2 \psi +
\cos \theta\,\sin 2\phi\,\cos2\psi\,,\nonumber \\
\label{eq:antennaPatterns}
\end{eqnarray}
where $\theta$ and $\phi$ are the polar and azimuth angles specifying the position of the binary in the detector frame, and $\psi$ is the polarization angle. Also, in Eq.~(\ref{eq:cqeq}),  
\begin{eqnarray}
C_+ (t) &=&\frac{1}{2}\cos^2\Theta(\sin^2\alpha-\cos^2i\cos^2\alpha)  \nonumber \\
&& +\frac{1}{2}(\cos^2i\sin^2\alpha -\cos^2\alpha) \nonumber \\
&& -\frac{1}{2}\sin^2\Theta\sin^2i -\frac{1}{4}\sin2\Theta\sin2i\cos\alpha\,, \nonumber \\
S_+ (t) &=& \frac{1}{2}(1+\cos^2\Theta)\cos i\sin2\alpha+\frac{1}{2}\sin2\Theta\sin i\sin\alpha\,, \nonumber \\
C_{\times} (t) &=& -\frac{1}{2}\cos\Theta(1+\cos^2i)\sin2\alpha-\frac{1}{2}\sin\Theta\sin2i\sin\alpha\,, \nonumber \\
S_{\times} (t) &=&-\cos\Theta\cos i\cos2\alpha-\sin\Theta\sin i\cos\alpha\,,
\label{steq}
\end{eqnarray}
where $\Theta$ is the angle between the line of sight from the detector to the source and the $z$-axis of the coordinate system (direction of initial $\bJ$). Note that, while $\Theta$ is a constant, the angles $\alpha$ and $i$ vary throughout the evolution of the binary (see Fig.~\ref{fig:SourceFrame}). 

The expression for the observed GW signal in a detector given in Eq.~(\ref{eq:cqsqwaveforms}) can be rewritten as 
\begin{equation}
h(t) = A(t) \cos \left[ \Psi(t) + \Psi_0 \right] ,
\end{equation}
where 
\begin{eqnarray}
\label{eq:hOfT2}
A(t) & = & \sqrt{C^2_Q(t) + S^2_Q(t)} \, , \nonumber \\
\Psi(t) & = & 2 \Phi(t) - \arctan \left[\frac{S_Q(t)}{C_Q(t)} \right] \,,
\end{eqnarray}
and $\Psi_0$ is a constant that depends on the initial configuration of the binary. Note that $S_Q(t)$ and $C_Q(t)$ are oscillatory in nature. These expressions show that the precession of the orbital plane is causing amplitude- and phase modulation in a GW signal $h(t)$ observed by a fixed detector. 

\subsection{Initial conditions and truncation conditions}
\label{sec:PrecIniCond} 

In order to simulate a GW signal observed in a detector, we have to specify, apart from the masses and spins, the initial orbital phase $\varphi_0$, two angles $\Theta$ and $\psi$ describing the initial orientation of the binary in the detector frame, and two angles $\theta$ and $\phi$ describing the sky-location of the binary in the detector frame. In this paper, we will assume that the binary is located optimally with respect to the detector: $\theta = \phi = 0$. Also, we set the initial phase $\varphi_0 = 0$. The evolution of the binary is started at the specified low-frequency (20 Hz) and truncated when $E'(v)/\Flux(v) = 0$~\footnote{Overlaps are computed considering only frequencies below the Schwarzschild ISCO frequency. Target waveforms are evolved until the highest possible frequency in order to minimize the in-band contamination due to the sharp truncation of the time-domain waveform (``edge effects'').}. 

Spins are specified in a coordinate system whose $z$-axis is determined by the initial Newtonian orbital angular momentum $\LNh$. After computing $\bJ \simeq \LN + \bS_1+\bS_2$, we rotate $\bS_1$, $\bS_2$ and $\LNh$ such that initial $\bJ$ is pointed along the $z$-axis. 

It is often convenient to describe the spin vectors $\bchi \equiv \{\chi_x, \chi_y, \chi_z\}$ in terms of a spin magnitude $||\bchi||$ and two spin angles $\theta_S$ and $\phi_S$:
\begin{eqnarray}
\chi_x & =& ||\bchi|| \, \sin \theta_S \, \cos \phi_S, \nonumber \\ 
\chi_y & =& ||\bchi|| \, \sin \theta_S \, \sin \phi_S, \nonumber \\ 
\chi_z & =& ||\bchi|| \, \cos \theta_S,
\end{eqnarray}
where $\theta_S$ is referred to as the spin tilt angle.

\section{Detection of binary inspiral signals}
\label{sec:Detection}

The GW signal $h(t; \blambda)$ observed in a detector is a function of the set of physical parameters $\blambda$ of the binary, such as the component masses and the spins. Detecting the signal requires analyzing (noisy) interferometric data. We assume that the detector noise $n(t)$ follows a zero-mean Gaussian distribution, characterized by its (one-sided) power spectral density (PSD) $S_h(f)$. We also assume the noise to be additive. This implies that when a signal is present in the data $d(t)$, then 
\be
d(t) = h(t; \blambda) + n(t) \,.
\ee
Under the above assumptions about the characteristics of detector noise, the Neyman-Pearson criterion \cite{Helstrom} leads to an optimal search statistic, which when maximized over the overall amplitude of the signal, is the cross-correlation of the data with a normalized template,
\be\label{crosscor}
\rho \equiv \langle \hat{h}(\blambda),d\rangle \ \ ,
\ee
where the normalized template is $\hat{h}(f) \equiv {\tilde h}(f)/\sqrt{\langle h,\>h \rangle} $. The angular bracket denotes the following noise-weighted inner product, called the \emph{overlap}: 
\be
\label{eq:innerprod}
\langle a ,\>b \rangle = 4 \Re \int_{\flow}^{\fcut} \! df\>{\tilde{a}^* (f) \,\tilde{b}(f) \over S_{h}(f)} \ \ , 
\ee
where $\tilde{a}(f)$ and $\tilde{b}(f)$ are the Fourier transforms of $a(t)$ and $b(t)$, respectively. Also, $\flow$ is a low-frequency cutoff of the detector noise and $\fcut$ is an upper cutoff frequency where the templates cease to describe the true signal with sufficient accuracy (e.g., due to the PN approximation breaking down).
The optimal SNR of the filter is given by 
\begin{equation}
\label{eq:optimSNR}
\rho_\mathrm{opt} \equiv \langle  h(\blambda),\>h(\blambda) \rangle ^{1/2}.
\end{equation}

In a ``blind'' search in detector data, where none of the binary's parameters are known \emph{a priori}, the search for a GW signal requires maximizing $\rho$ over a ``bank'' of templates corresponding to different values of $\blambda$. Apart from the physical parameters $\blambda$, the waveform also depends on the (unknown) time of arrival $t_0$ of the signal at the detector, and the corresponding phase $\Psi_0$. Maximization over $\Psi_0$ is effected by using two orthogonal templates for each combination of the physical parameters~\cite{schutz-91}, and the maximization over $t_0$ is attained efficiently with the help of the Fast Fourier Transform (FFT) algorithms. 

The inner product in Eq.(\ref{eq:innerprod}) between the template $x(\bLambda)$ (described by the set of parameters $\bLambda$) and the target signal $h(\blambda)$ maximized over $\Psi_0$ and $t_0$ is called the \emph{match}: 
\begin{equation}
\label{eq:matchDefn}
\mathrm{match} = \mathrm{max}_{\Psi_0,t_0} \, \langle  \hat{x}(\Psi_0,t_0,\bLambda),\>\hat{h}(\blambda) \rangle. 
\end{equation}
The fraction of optimal SNR  retrieved by a sub-optimal template bank is given by the match maximized over all the template parameters, called the \emph{fitting factor}~\cite{Apostolatos:1995pj}: 
\begin{equation}
\label{eq:FFDefn}
\mathrm{FF} = \mathrm{max}_{\Psi_0,t_0,\bLambda} \, \langle  \hat{x}(\Psi_0,t_0,\bLambda),\>\hat{h}(\blambda) \rangle. 
\end{equation}
Template families with high fitting factors (FF $\gtrsim$ 0.97) are considered to be ``effectual'' for signal detection, while templates with high values of match (match $\gtrsim$ 0.97) are considered to be both effectual in detection and ``faithful'' in estimating the parameters~\cite{DIS98}.

\subsection{Computing overlaps}

Computing the inner product in Eq.(\ref{eq:innerprod}) requires modeling the PSD of the detector noise. In this paper, we use a fit to the expected PSD of the ``zero-detuning, high power'' (zero-detuning of the signal recycling mirror, with full laser power) configuration~\cite{AdvLIGOPSD} of Advanced LIGO~\cite{0264-9381-27-8-084006}:
\begin{eqnarray}
S_h(f) & = & 10^{-48}\left(0.0152\,x^{-4} + 0.2935\,x^{9/4} + 2.7951\,x^{3/2} \right. \nonumber \\
       && -  \left. 6.5080\,x^{3/4} + 17.7622\right),
\end{eqnarray}
where $x = f/245.4$. The low-frequency cutoff (due to the elevated seismic noise) is assumed to be 20 Hz. 

The upper cutoff frequency of the integral in Eq.(\ref{eq:innerprod}) is chosen as the frequency of the innermost stable circular orbit (ISCO) of a test particle around a Schwarzschild black hole $f_\mathrm{ISCO} = v^3_\mathrm{ISCO}/(\pi m)$, where $v_\mathrm{ISCO} = 1/\sqrt{6}$. Note that, for the case of comparable-mass binaries, ISCO is a poorly defined quantity (see, e.g.~\cite{Favata:2010ic}), and the validity of the PN approximation can be only tested by comparing against fully general-relativistic numerical simulations (see~\cite{Buonanno:2006ui,PhysRevLett.99.181101,Hannam:2007ik,Gopakumar:2007vh,Boyle:2007ft,Hannam:2010ec,Campanelli:2008nk} for some work in this direction). Nevertheless, $f_\mathrm{ISCO}$  can be treated as a convenient cutoff frequency, especially since we will be computing overlaps with two different template families (spinning and non-spinning), and a fixed upper cutoff frequency will provide a fair comparison between the two. Indeed, in the case of spins (nearly) anti-aligned with the orbital angular momentum, the ``actual'' ISCO frequency could be significantly lower than $f_\mathrm{ISCO}$, and the time-domain waveforms discussed in Sec.~\ref{sec:targetWaves}  will terminate before $f_\mathrm{ISCO}$. This will reduce the overlaps of the templates with the target signals. But, since different template families will be affected in the same way, we neglect this effect. But we note that, in a actual search it might be more appropriate to use a spin-dependent cutoff frequency. 
Fourier transform of the time-domain target signals is computed with the help of the \texttt{FFTW} library~\cite{FFTW}. In order to avoid artifacts associated with the abrupt start and stop of the waveform in the time domain, a tapering window~\cite{McKechan:2010kp} is applied prior to computing the FFT. Maximization of the overlaps over the physical parameters (masses and spins) is performed with the aid of the Nelder-Mead downhill simplex algorithm \texttt{amoeba}~\cite{amoeba,NRecipes} with the implementation described in the Section III\,B of~\cite{Ajith:2009fz}, except that the mass parameters used in the maximization procedure are the chirp mass $\mathcal{M} \equiv m \, \eta^{3/5}$ and $\eta$.

\section{Fourier domain templates for inspiralling binaries with non-precessing spins}
\label{sec:EffSpinTempl}

In the case of binaries with spins aligned/anti-aligned with the orbital angular momentum, it can be seen from Eq.~(\ref{eq:precessionEqnsS}) that $d\bS_i/dt = 0$, and hence the spins and the angular momentum do not precess. This will considerably simplify the complexity of the GW signal, and all the spin effects can be described by two parameters $\chis \equiv \chisdL \equiv (\chi_1+\chi_2)/2$ and $\chia \equiv \chiadL \equiv (\chi_1-\chi_2)/2$, which remain constant throughout the evolution.  

Since the cross correlation between the data and the template is most efficiently computed in the Fourier domain by using the FFT, waveform templates in the Fourier domain are computationally cheaper. In the case of binaries with non-precessing spins, Fourier transform of the time-domain templates can be computed analytically using the stationary phase approximation~\cite{DIS01}, which can be written as 
\begin{equation}
\label{eq:SPAtempl}
\tilde{h}(f; \chi_1,\chi_2) \equiv A(f) \, e^{-\rmi \left[2\pi f t_0 + \Psi_0 + \Psi(f) - \pi/4 \right]}. 
\end{equation}
Above, 
\begin{equation}
A(f) = \mathcal{C} \, \frac{2 \eta m}{D} \, v_f^2 \left[\frac{dF(v_f)}{dt}\right]^{-1/2}
\label{eq:AofF1}
\end{equation}
where $v_f = (\pi m f)^{1/3}$, $F(v_f) = v_f^3/(\pi m)$ is the instantaneous GW frequency (of the dominant harmonic) evaluated at the stationary point $v_f$, $\mathcal{C}$ is a numerical constant that depends on the relative position and inclination of the binary with respect to the detector ($\mathcal{C} = 1$ for optimally located and oriented binaries), $t_0$ is the time of arrival of the signal at the detector, $\Psi_0$ the corresponding phase, and~\cite{Poisson:1995ef,Arun:2009} 
\begin{align}\label{eq:PsiofF}
\Psi(f) & = \frac{3} {128 \eta \, v^5} \left\{1 + v^2 \left[ \frac{55 \eta }{9}+\frac{3715}{756} \right] \right. \nonumber \\ 
    & \quad + v^3 \left[4\,\beta -16 \pi \right]   \nonumber\\
    & \quad + v^4 \left[\frac{3085 \eta ^2}{72}+\frac{27145 \eta }{504}+\frac{15293365}{508032} -10 \, \sigma \right]  \nonumber\\ 
    & \quad + v^5 \left[\frac{38645 \pi}{756}-\frac{65 \pi  \eta }{9} -\gamma \right] (3 \ln (v)+1)  \nonumber\\ 
    & \quad + v^6 \left[-\frac{6848 \gamma_E }{21}-\frac{127825 \eta ^3}{1296}+\frac{76055 \eta^2}{1728} \right.  \nonumber\\
    & \qquad \left. + \left(\frac{2255 \pi ^2}{12}-\frac{15737765635}{3048192}\right) \eta -\frac{640 \pi^2}{3} \right.  \nonumber\\
    & \qquad \left. +\frac{11583231236531}{4694215680}-\frac{6848 \ln (4v)}{21}\right]  \nonumber\\
    & \quad \left. + v^7 \left[-\frac{74045 \pi  \eta ^2}{756}+\frac{378515 \pi  \eta }{1512}+\frac{77096675 \pi }{254016} \right] \right\},
\end{align}
where the terms 
\begin{eqnarray}
\label{eq:BetaSigmaGammaDefn}
\beta & = & \frac{113}{12} \left(\chis + \delta \chia -\frac{76\eta }{113} \chis \right) , \nonumber \\ 
\sigma & = & \chia^2 \left(\frac{81}{16}-20 \eta \right)+\frac{81 \chia \chis \delta }{8}
    +\chis^2 \left(\frac{81}{16}-\frac{\eta}{4}\right), \nonumber \\ 
\gamma & = & \chia \delta  \left(\frac{140 \eta}{9}+\frac{732985}{2268}\right) 
    +\chis \left(\frac{732985}{2268} -\frac{24260 \eta }{81} -\frac{340 \eta^2}{9}\right), \nonumber \\
\end{eqnarray}
denote the leading-order spin-orbit coupling, leading-order spin-spin coupling, and next-to-leading-order spin-orbit coupling, respectively.

The time-derivative of the GW frequency appearing in Eq.(\ref{eq:AofF1}) can be written as: 
\begin{equation}
\frac{dF(v_f)}{dt} = \frac{dF}{dv_f} \frac{dv_f}{dE} \frac{dE}{dt} = \frac{3v_f^2}{\pi m^2} \left[\frac{-\Flux(v_f)}{E'(v_f)} \right],
\end{equation}
where we have used the energy balance equation $m\, dE/dt = -\Flux(v)$, and the definition $E'(v) \equiv dE/dv$. This can be plugged back to Eq.(\ref{eq:AofF1}) to get a closed form expression of $A(f)$. 
\begin{equation}
A(f) = \mathcal{C} \, \frac{2 \eta m^2 v_f}{D} \sqrt{\frac{\pi}{3}} \, \left[\frac{-E'(v_f)}{\Flux(v_f)} \right]^{1/2}. 
\end{equation}
Using the re-expansion of $E'(v)/\Flux(v)$ given in Eq.(\ref{eq:EbyF}) we get 
\begin{align}\label{eq:AofF2}
A(f) & = \mathcal{C}\,\frac{m^{5/6}}{D\,\pi^{2/3}} \left(\frac{5\eta}{24}\right)^{1/2}f^{-7/6} \left\{1 + v^2 \left[\frac{11 \eta }{8}+\frac{743}{672} \right] \right.\nonumber \\  
    &\quad + v^3 \left[\frac{\beta}{2} -2 \pi \right] \nonumber \\ 
    &\quad + v^4 \left[\frac{1379 \eta ^2}{1152} +\frac{18913 \eta }{16128}+\frac{7266251}{8128512} - \frac{\sigma}{2}\right] \nonumber\\ 
    &\quad + v^5 \left[\frac{57 \pi  \eta}{16}-\frac{4757 \pi }{1344} + \epsilon \right] \nonumber\\ 
    &\quad + v^6 \left[\frac{856 \gamma_E }{105}+\frac{67999 \eta ^3}{82944}-\frac{1041557 \eta
   ^2}{258048}-\frac{451 \pi ^2 \eta }{96}  \right. \nonumber\\
    &\qquad \left.  +\frac{10 \pi ^2}{3} +\frac{3526813753 \eta }{27869184}
        -\frac{29342493702821}{500716339200} \right. \nonumber \\ 
    &\qquad \left.    +\frac{856 \ln (4v)}{105} \right] \nonumber\\ 
    &\quad \left. + v^7 \left[-\frac{1349 \pi  \eta ^2}{24192}-\frac{72221 \pi  \eta }{24192}-\frac{5111593 \pi }{2709504} \right] \right\},
\end{align}
where 
\begin{equation}
\label{eq:EpsilonDefn}
\epsilon = \left(\frac{502429}{16128}-\frac{907 \eta }{192}\right) \delta \chia 
        +\left(\frac{5\eta ^2}{48}-\frac{73921 \eta }{2016} +\frac{502429}{16128}\right) \chis. 
\end{equation}
Note that the higher order terms in $A(f)$ are not due to the PN corrections to the (time-domain) amplitude of the waveform, but come from the higher order corrections to $dF/dt$. The standard practice followed in the literature is to truncate the expansion of $dF/dt$ in the leading order. But we found that including the higher order corrections to $dF/dt$ gives a better agreement with the exact (numerical) Fourier transform of the time-domain signals. 

It is well known that the leading order spin term $\beta$ (spin-orbit coupling) appearing at 1.5PN order in the amplitude and phase can be represented by a single parameter (see e.g.~\cite{Poisson:1995ef}), which we call the ``reduced spin'' parameter: 
\begin{equation}
\chi \equiv \chis + \delta \chia -\frac{76\eta }{113} \chis. 
\end{equation}
Note that this is different from the ``effective-spin'' $\bS_\mathrm{eff} \equiv (\chis + \delta \chia - \eta \,\chis/2)\,m^2$, introduced by Damour~\cite{PhysRevD.64.124013} to describe the leading order spin-orbit term in the conservative dynamics of the binary. In some sense, the reduced-spin parameter $\chi$ describes both the conservative and radiative dynamics of the binary as encoded in the GW signal. 

\begin{figure}[t]
\begin{center}
\includegraphics[width=2.5in]{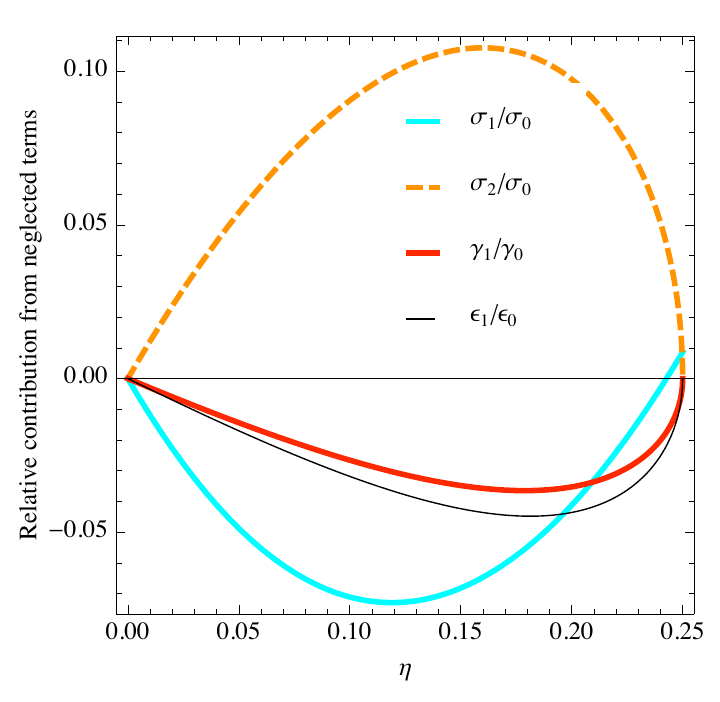}
\caption{Relative significance of the spin terms neglected in the ``reduced-spin'' templates, as a function of the symmetric mass ratio $\eta$. The plot shows the ratios of the coefficients of the spin terms $\chi_s$ and $\chi_a$ (which are neglected in ``reduced-spin'' templates) with the coefficients of $\chi$ (which are included). It can be seen that the relative contribution of the neglected terms is small ($\leq 10\%$), and the dominant spin effects are described by the single parameter $\chi$. See Eq.(\ref{eq:BetaSigmaGammaDefn}) for the definition of $\sigma_i, \gamma_i$ and $\epsilon_i$.}
\label{default}
\label{fig:NeglectedTermsRelSignificance}
\end{center}
\end{figure}

\begin{figure}[t]
\begin{center}
\includegraphics[width=3.4in]{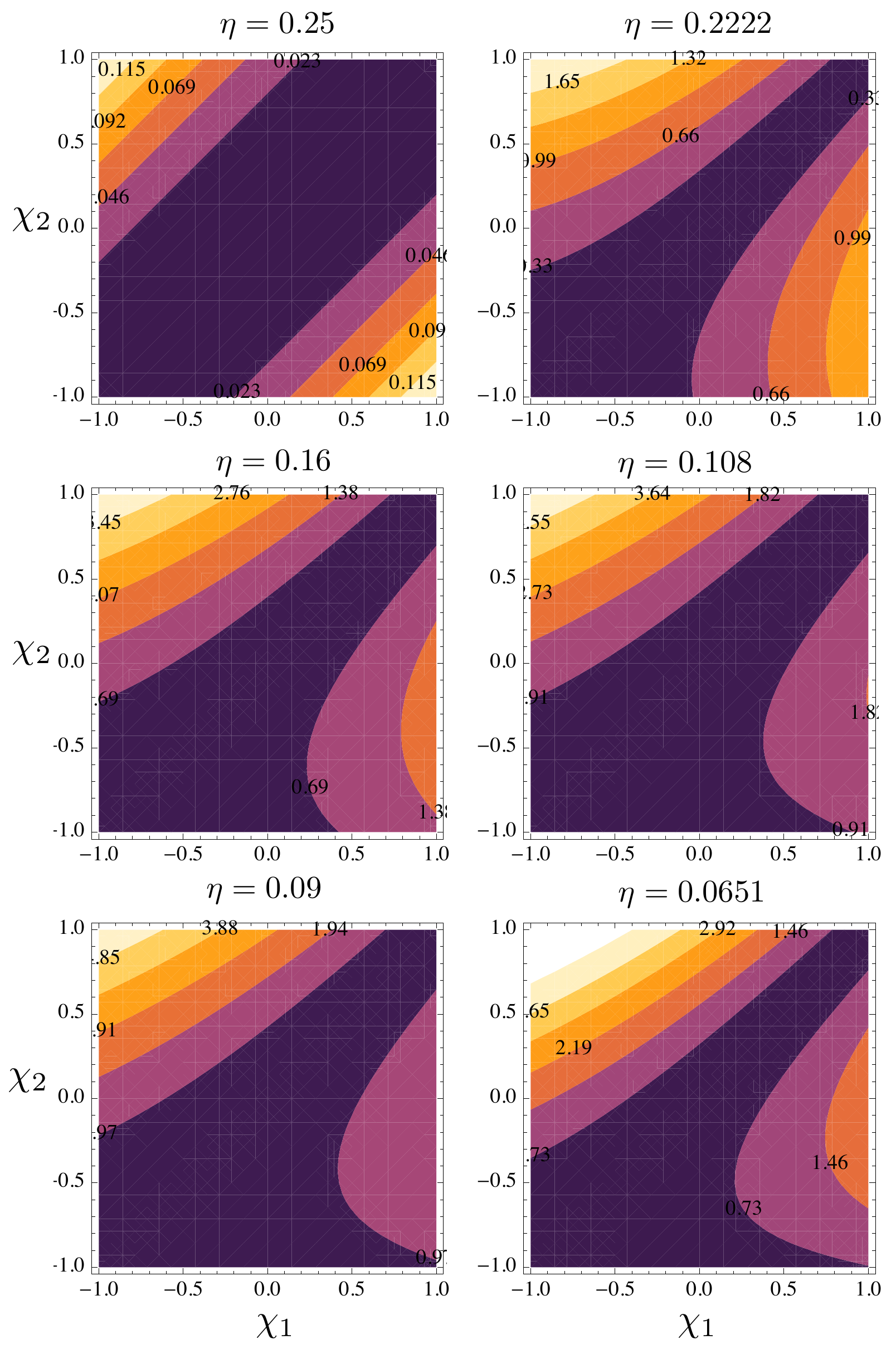}
\caption{Contribution to the phase $\Psi(f)$ at ISCO from the spin terms neglected in the ``reduced-spin'' template family. The plots show the contours of $|\Delta \Psi_\mathrm{ISCO}|$ (see Eq.~(\ref{eq:DeltaPsiISCO}) for definition) as a function of the spin parameters $\chi_1$ and $\chi_2$ of the two compact objects ($\chi_2$ being the spin of the more massive object), for the case of six different mass ratios. Darker shades correspond to lower values.}
\label{default}
\label{fig:Psi4sandPsi5SDiffContours}
\end{center}
\end{figure}

The spin-dependent terms $\beta, \sigma, \gamma$ and $\epsilon$ defined in Eqs.(\ref{eq:BetaSigmaGammaDefn}) and (\ref{eq:EpsilonDefn}) can be written in terms of the reduced-spin parameter $\chi$ as: 
\begin{eqnarray}
\label{eq:BetaSigmaGammaDefn}
\beta  & = & 113 \, \chi/12 , \nonumber \\ 
\sigma & = & \sigma_0 \, \chi^2 + \sigma_1 \, \chia^2 + \sigma_2 \, \chis \, \chia , \nonumber \\ 
\gamma & = & \gamma_0 \, \chi + \gamma_1 \, \chia                   , \nonumber \\
\epsilon & = & \epsilon_0 \, \chi + \epsilon_1 \, \chia,   
\end{eqnarray}
where 
\begin{eqnarray}
\label{eq:SigmaGammaEpsilonCoeffs}
\sigma_0 & = & -\frac{12769 \, (4 \eta -81)}{16 \, (76 \eta -113)^2} , \nonumber \\ 
\sigma_1 & = & \frac{\eta  \left(-115520 \eta ^2+359992 \eta -80569\right)}{(113-76 \eta )^2} , \nonumber \\ 
\sigma_2 & = & \frac{713 \, \delta  \eta }{76 \eta -113} , \nonumber \\ 
\gamma_0 & = & \frac{565 \left(17136 \eta ^2+135856 \eta -146597\right)}{2268 \, (76 \eta -113)} , \nonumber \\ 
\gamma_1 & = & -\frac{5 \, \delta  \eta \, (116676 \eta +417307)}{189 \, (76 \eta -113)} , \nonumber \\ 
\epsilon_0 & = & -\frac{113 \, \left(1680 \eta ^2-591368 \eta +502429\right)}{16128 \, (76 \eta -113)} , \nonumber \\ 
\epsilon_1 & = & -\frac{\delta  \eta \,  (116676 \eta +417307)}{336 (76 \eta -113)} . 
\end{eqnarray}
Figure~\ref{fig:NeglectedTermsRelSignificance} illustrates the fact that the coefficients ($\sigma_1, \sigma_2, \gamma_1, \epsilon_1$) of $\chi_a$ and $\chi_s$ are significantly smaller than the coefficients ($\sigma_0, \gamma_0, \epsilon_0$) of $\chi$. This gives the indication that the dominant spin effects in the waveform could be captured by just using the reduced spin parameter $\chi$. A template family solely described by the reduced spin parameter $\chi$, apart from $m$ and $\eta$, can be constructed by setting $\sigma_1 = \sigma_2 = \gamma_1 = \epsilon_1 = 0$ in Eqs.(\ref{eq:SPAtempl}) -- (\ref{eq:SigmaGammaEpsilonCoeffs}). i.e., 
\begin{equation}
\tilde{h}(f; \chi) = \tilde{h}(f; \chi_1, \chi_2),~~ \mathrm{with}~ \sigma_1 = \sigma_2 = \gamma_1 = \epsilon_1 = 0
\end{equation}
It is possible to get a rough idea of the effect of the neglected spin terms $\sigma_1, \sigma_2, \gamma_1, \epsilon_1$ at different regions in the parameter space by looking at their contribution to the Fourier domain phase at the ISCO frequency:  
\begin{align}
\label{eq:DeltaPsiISCO}
\Delta \Psi_\mathrm{ISCO} & \equiv \frac{-3}{128\eta} \, \Bigg[\frac{10 \, (\sigma_1 \, \chia^2 + \sigma_2 \, \chis \chia)}{ v_\mathrm{ISCO}} \nonumber \\
  & \qquad +  \gamma_1 \, \chia \left(1+ 3 \ln (v_\mathrm{ISCO})\right) \Bigg].
\end{align}
In Fig.~\ref{fig:Psi4sandPsi5SDiffContours}, we plot the contours of $|\Delta \Psi_\mathrm{ISCO}|$, which suggests that, for the equal-mass ($\eta = 0.25$) case there is little loss in neglecting the additional spin terms $\sigma_1, \sigma_2, \gamma_1$ and $\epsilon_1$ over most regions of the spin parameter space; but the differences get larger for the case of highly unequal masses. Nevertheless, it is likely that a template family described by one reduced spin parameter $\chi$ will be able to capture the more general family of signals (described by two spin parameters $\chis$ and $\chia$) at the cost of a parameter bias in $\chi$. 

\begin{figure*}[tb]
\begin{center}
\includegraphics[width=6.0in]{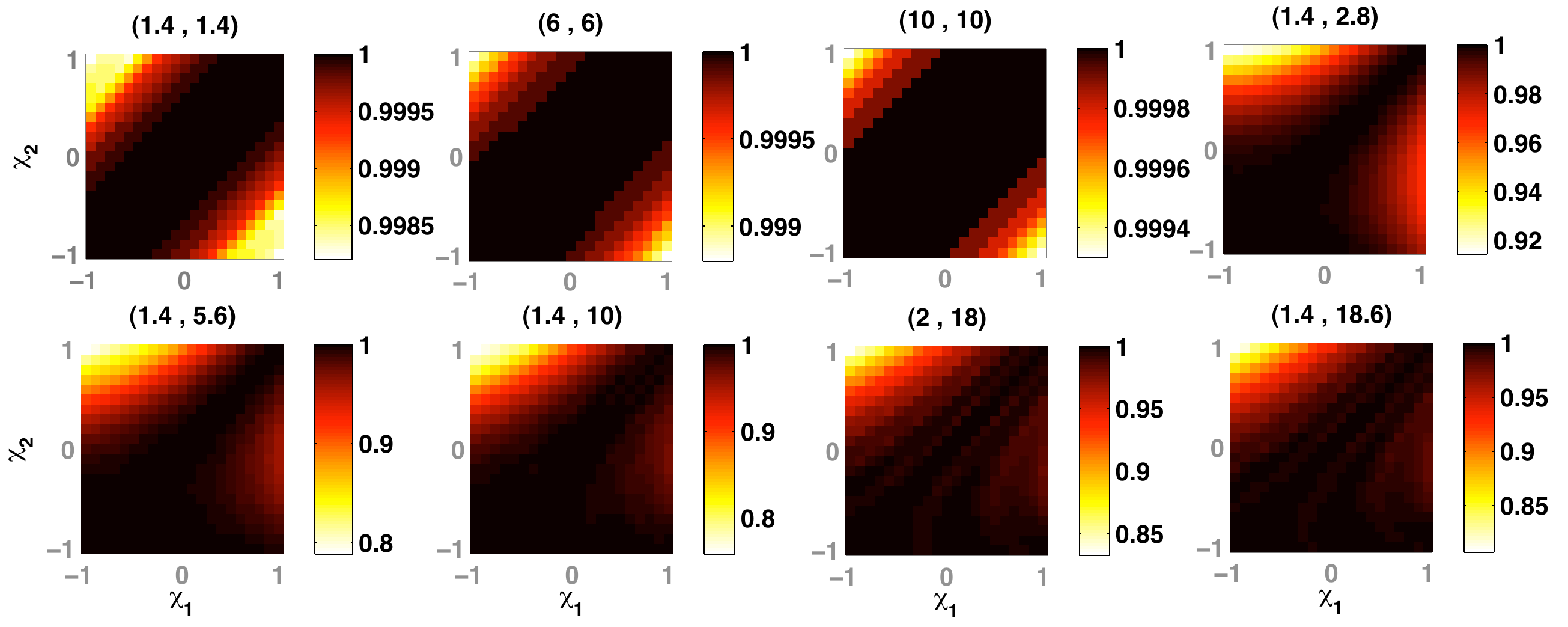}
\caption{Match of the reduced-spin template family $\tilde{h}(f; \chi)$ with non-precessing-spin signals $\tilde{h}(f; \chi_1, \chi_2)$ in Advanced LIGO noise spectrum. Horizontal axes report the spin $\chi_1$ of the lighter object and vertical axes report the spin $\chi_2$ of the more massive object. The title of each plot reports the component masses $(m_1,m_2)$ in units of $M_\odot$.}
\label{fig:TF2TF2ReduceMatch}
\end{center}
\end{figure*}

\begin{figure*}[tb]
\begin{center}
\includegraphics[width=6.0in]{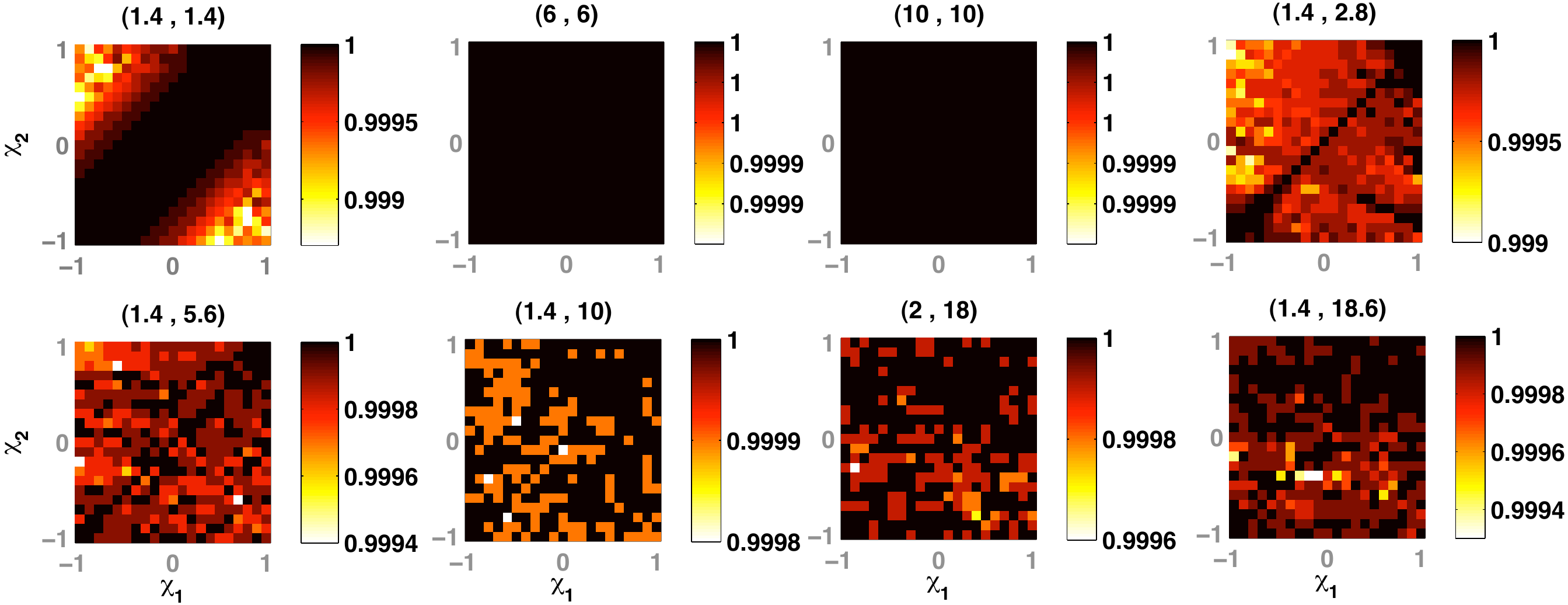}
\caption{Fitting factor (match maximized over $M$, $\eta$ and $\chi$) of the reduced-spin template family $\tilde{h}(f; \chi)$ with non-precessing-spin signals $\tilde{h}(f; \chi_1, \chi_2)$.
See Fig.~\ref{fig:TF2TF2ReduceMatch} for a full description.}
\label{fig:TF2TF2ReduceFF}
\end{center}
\end{figure*}

\begin{figure*}[tb]
\begin{center}
\includegraphics[width=6.0in]{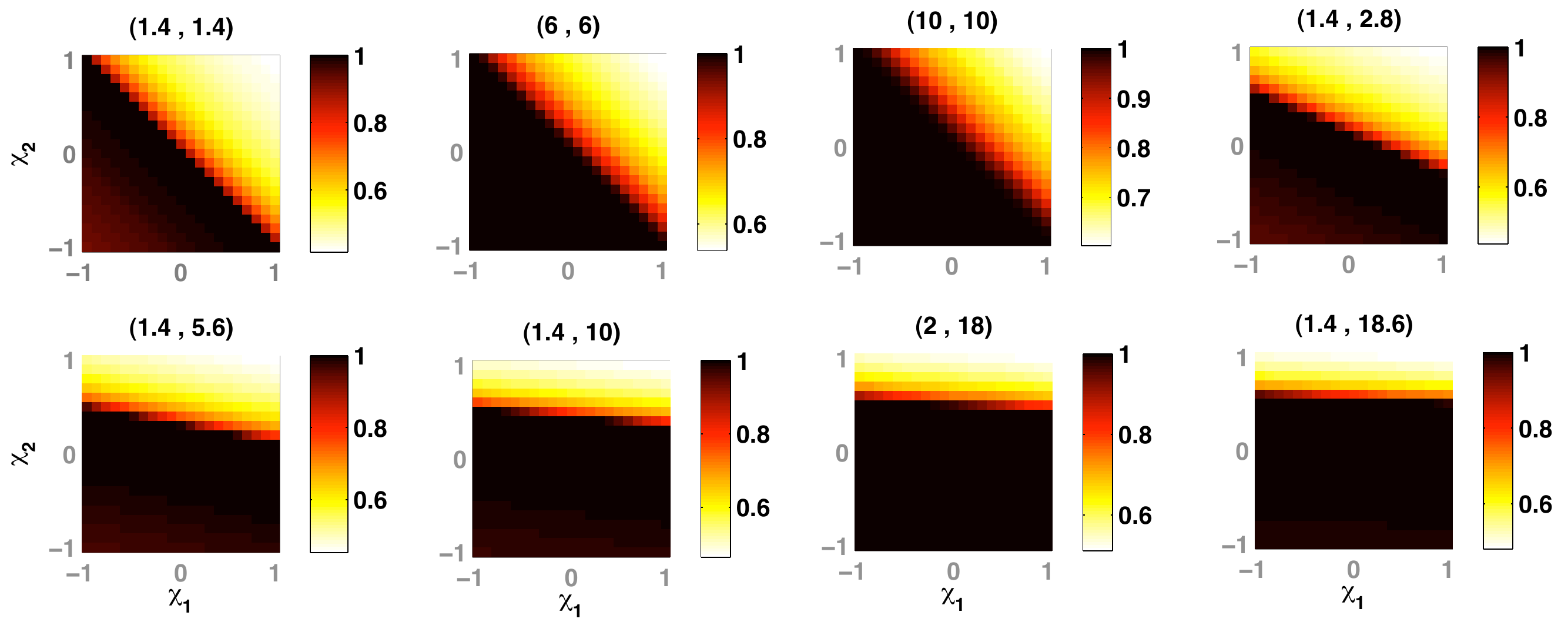}
\caption{Fitting factor (match maximized over $M$, $\eta$) of the \emph{non-spinning} PN template family $\tilde{h}(f; \chi_1=\chi_2=0)$ with non-precessing-spin signals $\tilde{h}(f; \chi_1, \chi_2)$. 
See Fig.~\ref{fig:TF2TF2ReduceMatch} for a full description.}
\label{fig:TF2TF2NSFF}
\end{center}
\end{figure*}

\begin{figure}[tb]
\begin{center}
\includegraphics[width=3.0in]{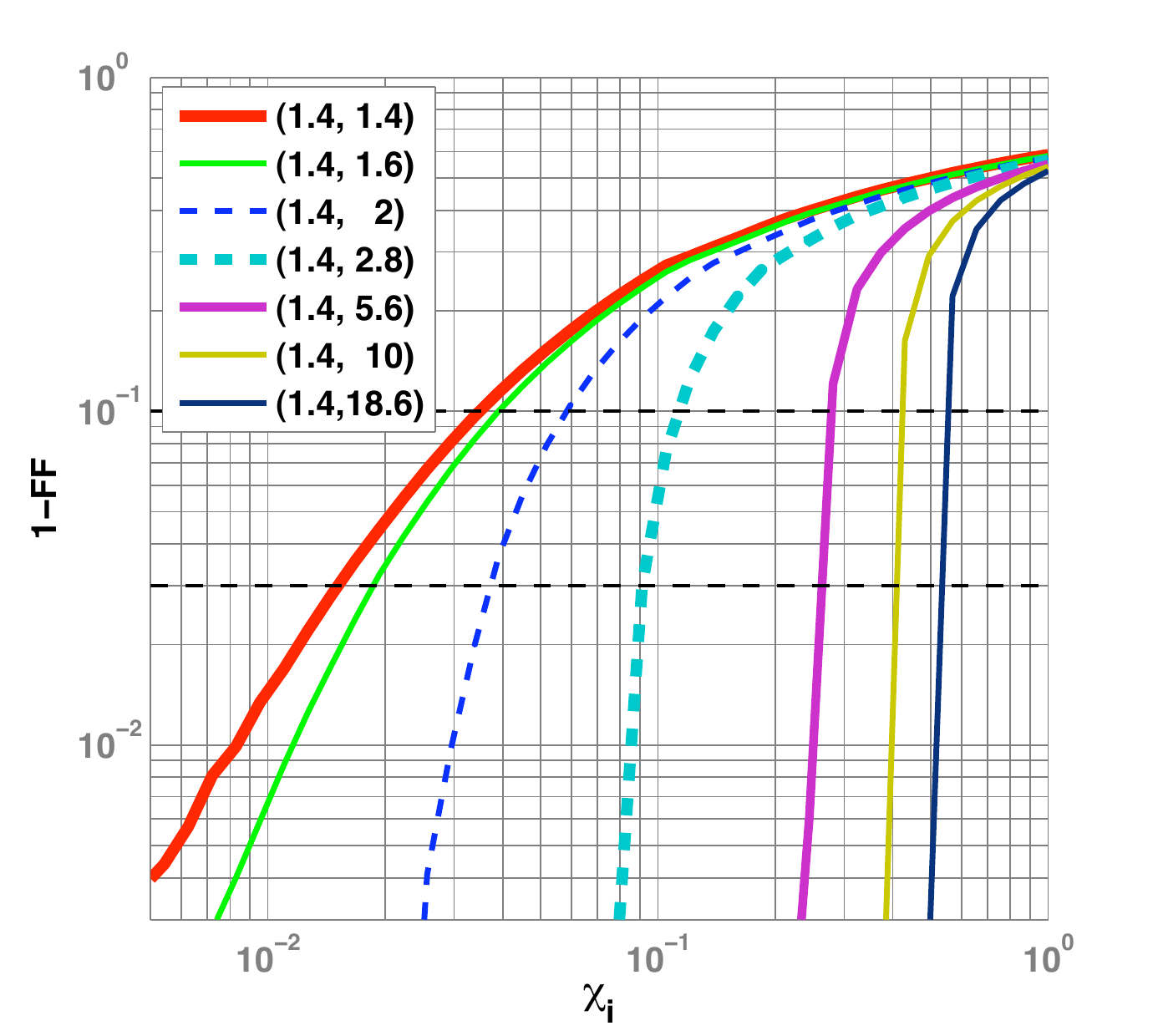}
\caption{Mismatch (1-FF) of the \emph{non-spinning} PN template family with \emph{non-precessing-spin} signals with equal spins. The horizontal axis reports the spins of the target binary ($\chi_1 = \chi_2$). Component masses of the target binaries (in units of $M_\odot$) are shown in the legend. 3\% and 10\% mismatch values are indicated by horizontal dashed (black) lines.} 
\label{fig:FF_TaylorF2SA_TaylorF2NS_Vs_Spin}
\end{center}
\end{figure}

\begin{figure*}[tb]
\begin{center}
\includegraphics[width=6.0in]{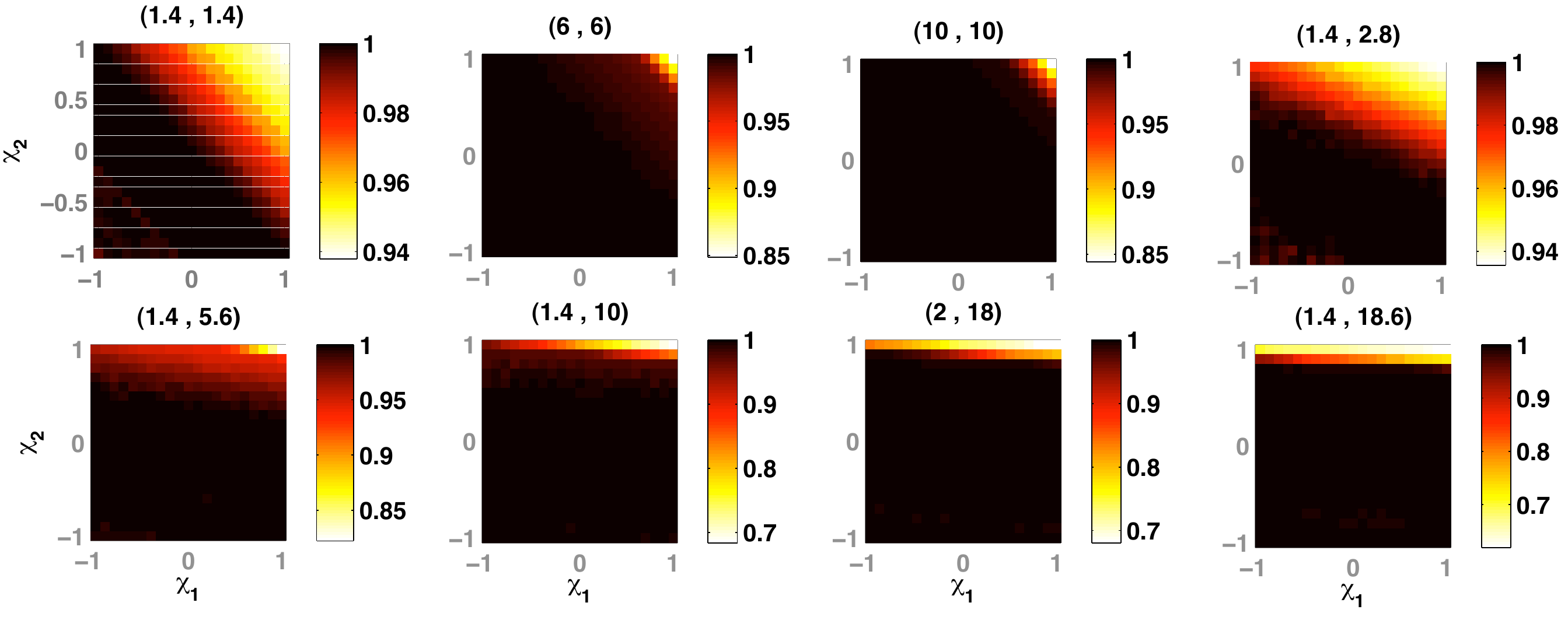}
\caption{Fitting factor (match maximized over $M$, $\eta$, $\beta$) of the PN template family considering only the leading-order spin effect (described by $\beta$) with non-precessing-spin signals in Advanced LIGO. 
See Fig.~\ref{fig:TF2TF2ReduceMatch} for a full description.}
\label{fig:FF_TaylorF2SA_TaylorF2LeadingSpin}
\end{center}
\end{figure*}


Figure~\ref{fig:TF2TF2ReduceMatch} shows the \emph{match} (see Eq.~\ref{eq:matchDefn} for definition) of the ``reduced-spin'' template family $\tilde{h}(f; \chi)$ with $\tilde{h}(f; \chi_1, \chi_2)$ for different values of the mass parameters $m_1$ and $m_2$ and spin parameters $\chi_1$ and $\chi_2$. It can be seen that the plots follow the same trends shown by Fig.~\ref{fig:Psi4sandPsi5SDiffContours} -- the ``reduced-spin'' template family produces very high matches ($> 0.99$) either when the masses are equal \emph{or} when the spins are equal. On the other hand, for the case of highly unequal masses \emph{and} spins, the match can be as low as 0.8. Figure~\ref{fig:TF2TF2ReduceFF} plots the fitting factor (see Eq.~\ref{eq:FFDefn} for definition) of the reduced spin template family. Fitting factor is greater than 0.999 over the entire parameter space, including extreme spins. This suggests that the reduced spin template family can be used in the detection of non-precessing-spin binaries with arbitrary mass ratios and spins \emph{causing no appreciable loss of SNR}. 

Figure~\ref{fig:TF2TF2NSFF} plots the fitting factor of the \emph{non-spinning} PN template family, constructed by setting $\beta = \sigma = \gamma = 0$ in Eqs.~(\ref{eq:SPAtempl})--(\ref{eq:BetaSigmaGammaDefn}), in detecting non-precessing signals $\tilde{h}(f; \chi_1, \chi_2)$, which demonstrates the effect of neglecting the spin terms. The biggest effect is in the case of binary neutron stars ($m_1 = m_2 = 1.4 M_\odot)$. Since the secular spin effects accumulate over the large number of GW cycles present in the detector band, even moderate spins can cause appreciable mismatches in the case of binary-neutron-star signals in Advanced LIGO. Note that these effects are fully captured by the reduced spin templates. The non-spinning template family is least efficient towards binaries with the spin of the more massive object \emph{aligned} (as opposed to \emph{anti-aligned}) to the orbital angular momentum. 

Figure~\ref{fig:FF_TaylorF2SA_TaylorF2NS_Vs_Spin} plots the \emph{mismatches} (1-fitting factor) of the non-spinning template family with non-precessing signals in the case of binaries with spins aligned to the angular momentum, as function of the spin of the target binary. Mismatches are computed for binaries with different component masses. For simplicity, the binary components are assumed to have equal spins ($\chi_1 = \chi_2$). It can be seen that, for the case of binary neutron stars ($m_{i} \leq 2 M_\odot$), even moderate spins ($\chi_i \simeq 0.015 - 0.1$) can cause appreciable mismatches (3\% -- 25\%). This is contrary to the prevailing assumption that the neutron star spins may not be relevant for GW detection. 

Finally, let us note that the ``reduced-spin'' template family proposed in this paper is different from the frequency-domain template family considering only the leading order spin-orbit term $\beta$. In reduced-spin templates, we are re-expressing the higher order spin terms (appearing at 2PN and 2.5PN order in amplitude and phase) in terms of the leading-order spin term in an approximate way, without increasing the dimensionality of the template bank. Indeed, the main reason for the improved effectualness of reduced-spin templates (as compared to non-spinning templates) is the inclusion of the leading order spin-orbit term. Nevertheless, inclusion of the higher order spin terms does provide some additional increase in effectualness. As an illustration, we plot in Figure~\ref{fig:FF_TaylorF2SA_TaylorF2LeadingSpin}, the fitting factor of the frequency-domain template family considering only the leading order spin term (but considering non-spinning terms up to 3.5PN). A comparison with Figure~\ref{fig:TF2TF2ReduceFF} will demonstrate the advantage of the reduced spin template family over the template family considering only the leading order spin effects.  

\section{Effectualness of the reduced-spin template family in detecting generic precessing binaries}
\label{sec:Precession}

\begin{table}[htdp]
\caption{Parameters used for the Monte-Carlo simulations of precessing PN binaries.}
\begin{center}
\begin{tabular}{ccccc}
\hline
\hline 
Parameter &&\vline&& Value \\
\hline 
$||\bchi_i|| ~ (m_i \geq 2 M_\odot)$ &&\vline&& uniform(0, 0.98) \\
$||\bchi_i|| ~ (m_i < 2 M_\odot)$    &&\vline&& uniform(0, 0.7) \\
$\theta_S$                           &&\vline&& uniform$(0, \pi)$ \\
$\phi_S$                             &&\vline&& uniform$(0, 2\pi)$ \\
$\Theta,\psi$                        &&\vline&& uniform$(0, \pi)$ \\
$\theta,\phi$                        &&\vline&& 0 \\
$N_\mathrm{sims}$                    &&\vline&& $\sim$ 2500 --- 5000 \\
\hline
\hline 
\end{tabular}
\end{center}
\label{tab:MonteCarloParams}
\end{table}%

\begin{figure}[tbh]
\begin{center}
\includegraphics[width=3.3in]{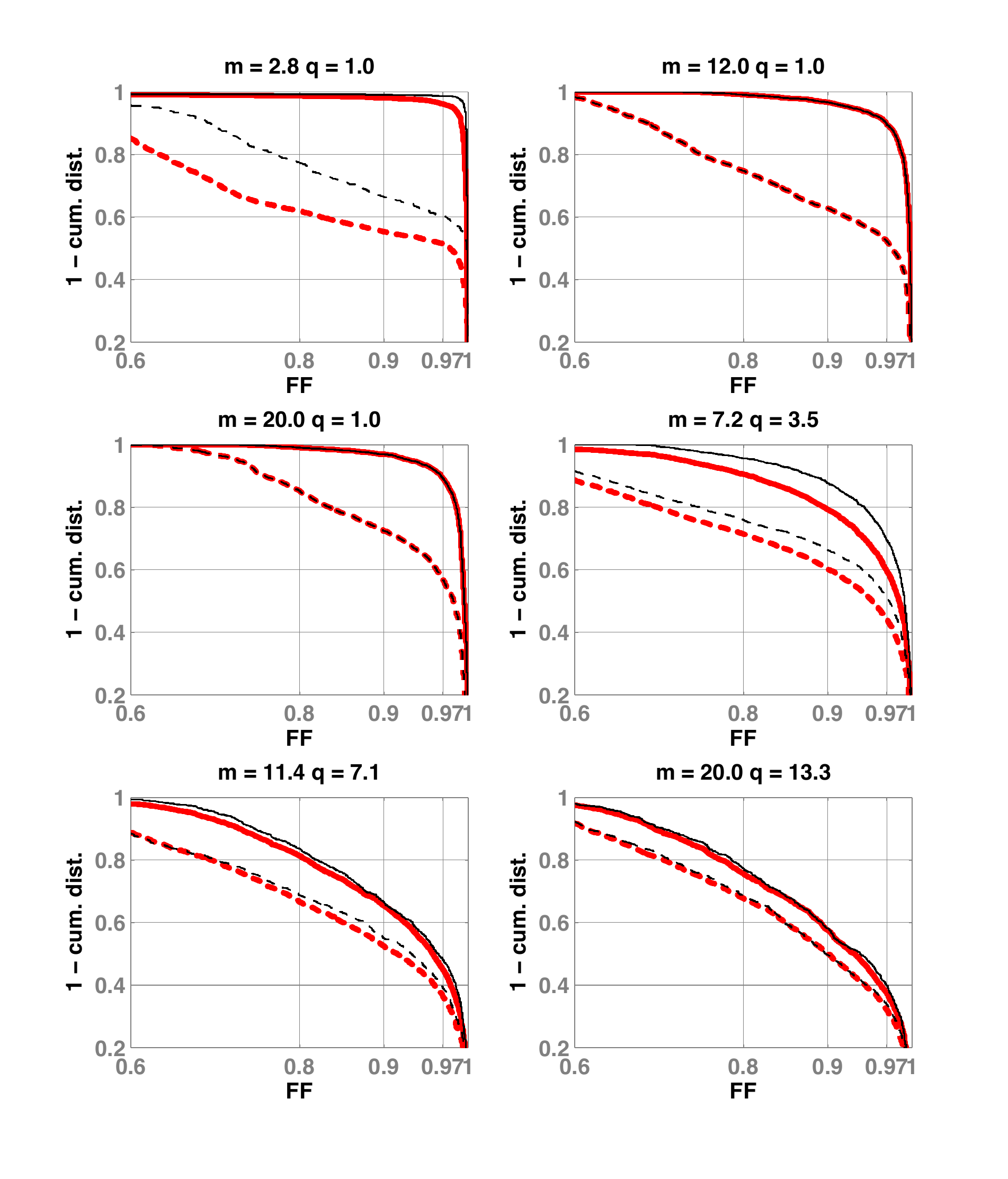}
\caption{Fraction of precessing PN binaries producing fitting factor $\geq$ FF with the reduced spin PN templates (thick red solid lines) and non-spinning PN templates (thick red dashed lines). The total mass $m/M_\odot$  and mass ratio $q$ are shown in the titles. Thin black lines plot the same except that the spin distribution of neutron stars is restricted to the interval (0, 0.3).}
\label{fig:FFHists}
\end{center}
\end{figure}

\begin{figure*}[tbh]
\begin{center}
\includegraphics[width=6.5in]{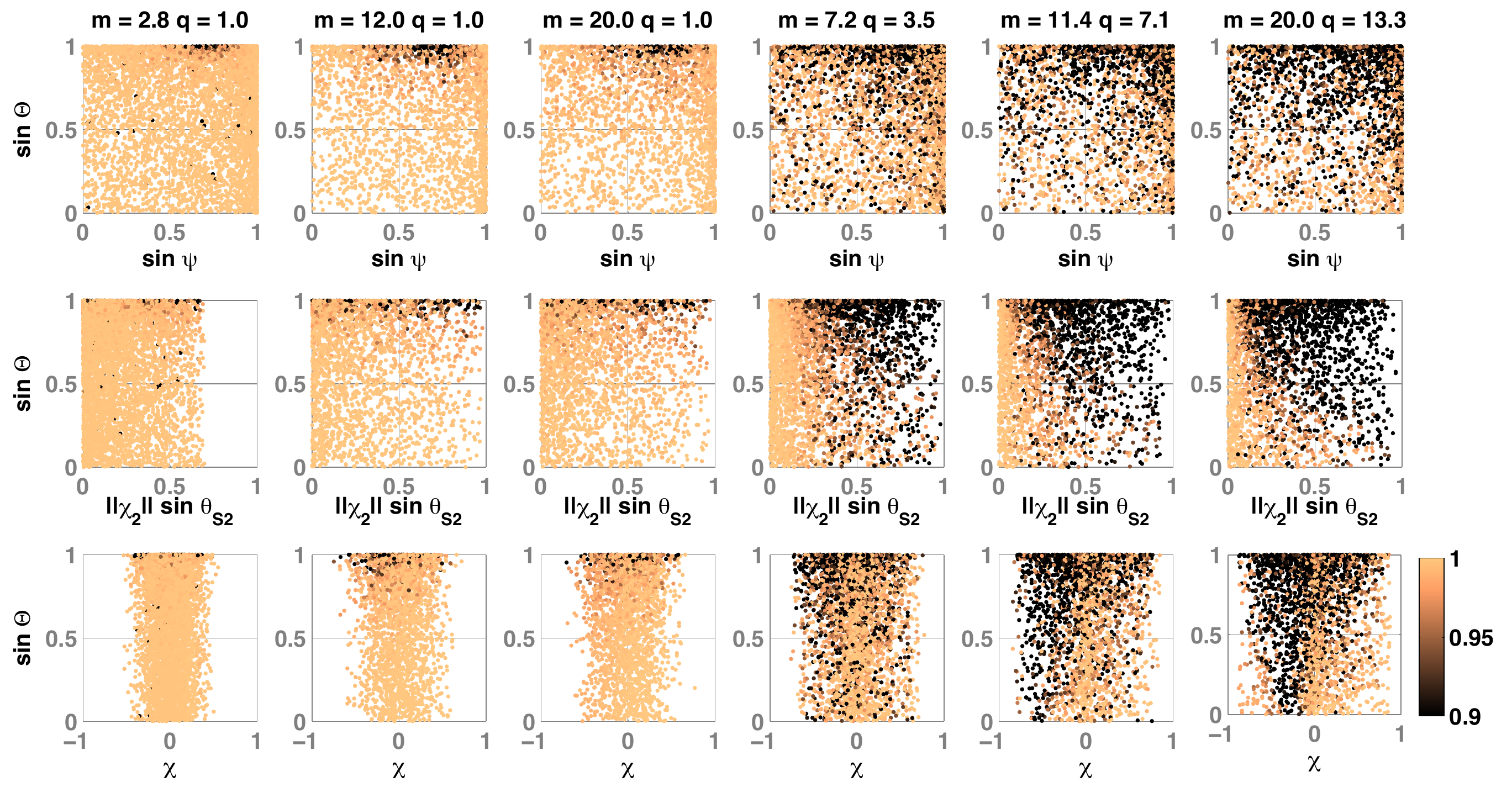}
\caption{Scatter plots of the fitting factor (indicated by the color of the dots) of the reduced spin template family with precessing PN binaries as a function of different parameters describing the initial configuration of the binary. The angles $\Theta$ and $\psi$ describe the initial orientation of the binary with respect to the detector, $||\chi_2||$ is the spin magnitude of the more massive compact object and $\theta_{S2}$ describes its orientation with respect to the orbital angular momentum (see Sec.~\ref{sec:PrecIniCond} for a complete description) while $\chi \equiv (1-76\eta/113) \, \chisdL + \delta \chiadL$ is the initial value of the reduced spin parameter. The total mass $m/M_\odot$ and mass ratio $q$ of the target population in each column are shown in the title.}
\label{fig:PrecFFScatterPlotsEffSpin}
\end{center}
\end{figure*}

\begin{figure*}[tbh]
\begin{center}
\includegraphics[width=6.5in]{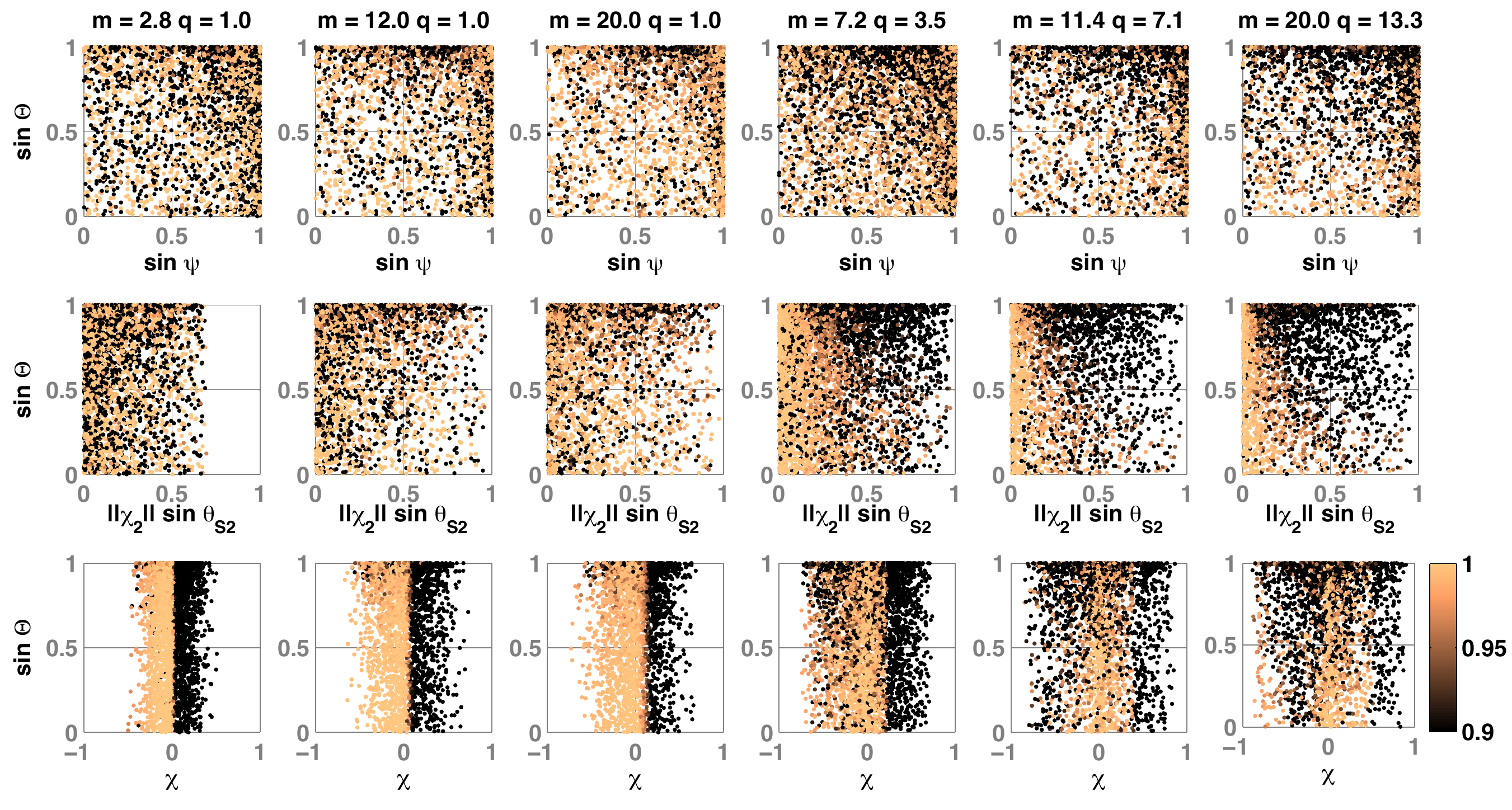}
\caption{Same as Fig.~\ref{fig:PrecFFScatterPlotsEffSpin}, except that here the fitting factors are calculated with respect to non-spinning PN templates}. 
\label{fig:PrecFFScatterPlotsNS}
\end{center}
\end{figure*}

In this section, we investigate the effectualness of the reduced-spin template family in detecting generic precessing PN binaries. There is suggestive indication that we will be able to model the most important spin effects in the orbital phase $\varphi(v)$ using the reduced spin parameter $\chi \equiv (1-76\eta/113) \, \chisdL + \delta \chiadL$ --- the leading-order spin orbit coupling in $\varphi(v)$ and $dv/dt$ is completely described by this parameter (see Eqs.~(\ref{eq:EbyF}), (\ref{eq:phiOfV}) and (\ref{eq:phasingformula2})). Indeed, for the case of precessing binaries, the angles between the spin and orbital angular momentum vectors change during the evolution, and hence the value of $\chi$. But it can be shown that, over a significant fraction of the parameter space that we are interested in, $\chi$ is a slowly evolving quantity. 

Let us consider the precession equations (\ref{eq:precessionEqnsS}) and (\ref{eq:Omega}). Considering only the leading order terms, the change in $\chi$ can be computed as: 
\begin{equation}
\frac{d\chi}{dt} \simeq -\frac{39 \, \delta }{226 \, m^5 \, \eta } \,v^6 ~ \bS_2 \, . \,(\bS_1 \times \LNh)
\end{equation}
It is evident that $\chi$ (in the leading order) remains a constant throughout the evolution either when one of the spins are zero ($\bS_i \rightarrow 0$) or when the spins are parallel to each other or to the orbital angular momentum ($\bS_2 \, . \, (\bS_1 \times \LNh) \rightarrow 0$), or when the masses are equal ($\delta \rightarrow 0$). Additionally, the change in $\chi$ is of the order of $v^6$, and hence smaller than the change in the spin vectors, which is of the order of $v^5$ (see Eqs.~(\ref{eq:precessionEqnsS}) and (\ref{eq:Omega}))~\footnote{We note that Galley \etal~\cite{Galley:2010rc} have identified a different combination of the spin scalar products as a (nearly) conserved quantity through the evolution.}. 

This observation \emph{suggests} that the reduced-spin template family described in Section~\ref{sec:EffSpinTempl} might be able to detect a large fraction of precessing binaries in the comparable mass regime. We expect the effectualness of the template family to deteriorate for the case of highly unequal masses. Also note that the precession of the orbital plane (described by the normal vector $\LNh$)  for a given spin configuration is inversely proportional to the symmetric mass ratio, as seen in Eq.~(\ref{eq:precessionEqnsL}) by taking the leading order terms, which also suggests that precessional effects (of the orbital plane) are minimal in the case of equal-mass binaries. Indeed, the GW signal observed by a detector contains additional modulations due to the precession of the orbital plane, as described by Eqs.~(\ref{eq:dPhiByDt}) and (\ref{eq:hOfT2}), which depend on the orientation angles $\Theta$ and $\psi$ of the binary apart from the spin magnitudes, spin orientations and the masses.   

In order to quantify the effectualness of the reduced-spin template family in detecting precessing binaries, we perform a Monte-Carlo simulation where we generate generic spinning binaries with random spins and compute the fitting factor of the reduced spin template family with these target signals. For binaries with $m \geq 2 M_\odot$, spin magnitudes $||\bchi_i||$ are randomly drawn from a uniform distribution in the interval (0, 0.98), and for binaries with $m < 2 M_\odot$ spin magnitudes are distributed in the interval (0,0.7).  The spin angles $\theta_S$ and $\phi_S$ are uniformly distributed in the intervals $(0, \pi)$ and $(0, 2\pi)$, respectively. Additionally, the inclination $\Theta$ of the initial total angular momentum vector $\bJ$ with respect to the line of sight from the detector, and the polarization angle $\psi$ are also uniformly distributed in the interval $(0, \pi)$. In the simulations, the binaries are assumed to be optimally located in the sky $(\theta = \phi = 0)$. Around 2500---5000 simulations were performed for each mass configuration. 
The simulation parameters are summarized in Table~\ref{tab:MonteCarloParams}. 

Figure~\ref{fig:FFHists} presents cumulative histograms showing the fraction of binaries producing fitting factors $\geq$ FF with the reduced-spin PN templates (solid lines) and non-spinning PN templates (dashed lines). These plots show that almost the entire population of equal-mass binaries can be detected using the reduced spin templates with fitting factors $\geq$ 0.97~\footnote{The small fraction of binaries with fitting factor $<$ 0.97 correspond to the orientations of the binary producing little or no observed signal in the detector -- see Fig.~\ref{fig:PrecFFScatterPlotsEffSpin} and the associated discussion in the text.}. On the other hand, only $\sim 51\% \, (52\%) \, 57\%$ of the equal-mass population with total mass $2.8 M_\odot \,(12 M_\odot)\, 20 M_\odot$ produces the same fitting factor with non-spinning PN templates, demonstrating the importance of taking into account the secular spin-dependent phase evolution of the binary. The loss of non-spinning templates is the highest for the case of the binary neutron star system $(1.4 M_\odot, 1.4 M_\odot)$, contradicting the expectation that spin effects may not be significant for the search for binary neutron stars. The effectualness of the non-spinning templates gradually improves with increase in total mass, owing to the smaller number of GW cycles present in the detector band. As expected, the effectualness of the reduced-spin templates decreases with increasing mass ratio $q$ ($\eta \ll 0.25$), since for binaries with significantly unequal masses, oscillatory effects of precession become important. Still, the reduced-spin template family performs considerably better than non-spinning templates: $\sim$ 60\% (45\%) 37\% of the population with $m_2/m_1 = 3.5\, (7.1) \, 13.3$ produces fitting factor $>$ 0.97 with reduced spin templates, while $\sim$ 44\% (36\%) 32\% produces the same fitting factor with non-spinning templates. For highly unequal-mass binaries, precessional effects are almost entirely determined by the spin of the larger compact object, and hence a template family describing precessional effects assuming only one spinning compact object, such as the physical template family proposed by BCV~\cite{BCV2,Pan:2003qt,FaziThesis,Harry:2010fr} should be able to model these binaries accurately.    

It is worthwhile to identify regions of the parameter space where the reduced spin template family is not effectual so that a more complex template family could be used in those regions. Figure~\ref{fig:PrecFFScatterPlotsEffSpin} shows scatter plots of the fitting factor of the reduced spin templates as a function of various parameters describing the initial configuration of the precessing binaries, while Fig.~\ref{fig:PrecFFScatterPlotsNS} shows the same for the case of non-spinning templates. The main conclusions one can draw from these figures are: 

\begin{enumerate}

\item In the case of equal-mass binaries, all the binaries producing fitting factor $<0.9$ with the reduced spin templates correspond to the initial orientations of $\Theta \simeq \pi/2$ and $\psi \simeq \pi/4$, producing little or no observed signal in the detector. 

\item In the case of unequal-mass binaries, most of the binaries producing fitting factor $<0.9$ with the reduced spin templates are significantly tilted with respect to the detector ($\Theta \gg 0$) and have spins of the more massive object highly non-aligned with the orbital angular momentum ($||\chi_2|| \sin \theta_{S2} \gg 0$). If there are strong astrophysical priors restricting the inclination angles or spin tilt angles of the target population to small values, the effectualness of this template family will be better than that presented in Fig.~\ref{fig:FFHists}. A possible scenario where this could apply is the GW followups of short-hard gamma ray bursts, where the opening angle of the jet is constrained to a few tens of degrees (see, e.g.,~\cite{Goldstein:2011ju})~\footnote{On the other hand, using GW observations we may want to test the accuracy of these astrophysical priors. In that case, the GW searches need to be as ``open-minded'' as possible.}.

\item For the case of non-spinning templates (Fig.~\ref{fig:PrecFFScatterPlotsNS}), the initial value of the reduced spin parameter $\chi \equiv (1-76\eta)/113 \, \chisdL + \delta \chiadL$ clearly separates the population of binaries producing poor fitting factor from those producing high fitting factor. In particular, in the case of equal-mass binaries, those with $\chi > 0$ have particularly poor fitting factor. This is consistent with the results we have seen for the case of non-precessing binaries, as shown in Fig.~\ref{fig:TF2TF2NSFF}, suggesting that the poor performance of the non-spinning templates is due to neglecting the secular spin-dependent effects in the phase evolution. For the case of binary neutron stars, as much as 25\% of the SNR can be lost if $\chi \simeq 0.1$. As the mass ratio increases ($\eta \ll 0.25$) other precessional effects also start to affect the performance of the template family. 

\end{enumerate}

\section{Conclusions}

In this paper, we presented a PN detection template family of gravitational waveforms from inspiralling compact binaries with non-precessing spins. The waveforms are reparametrized in such a way that all spin-dependent terms are described by a single ``reduced-spin'' parameter in an approximate fashion. We have shown that the template family has very high overlaps with non-precessing binary signals with arbitrary spins and mass ratios. The family is also effectual for the detection of a significant fraction of precessing binaries in the comparable-mass regime. This is due to the fact that, in the comparable-mass regime, the secular spin-dependent effects dominate the phase evolution (not the oscillatory effects), which are captured by the non-precessing templates. Since non-spinning templates neglect this effect, a significant fraction of SNR can be lost if non-spinning templates are used in the search for spinning binaries. 

The simple, closed-form expression of this frequency domain template family makes it very easy to implement this in search pipelines, and will provide considerable improvement over the non-spinning templates. Since a parameter-space metric~\cite{Owen:1995tm} can be easily worked out for this waveform family, the current techniques of template placement and multi-detector coincidence tests~\cite{Robinson:2008un} can be readily used in such a search. The parameterization of the inspiral waveforms using the reduced spin parameter $\chi$ can also be extended to the construction of inspiral-merger-ringdown waveform templates for binary black holes, such as the ones presented in~\cite{Ajith:2009bn,Santamaria:2010yb}.  

\smallskip 

\acknowledgments 
I would like to thank K. G. Arun, Duncan Brown, Alessandra Buonanno, Yanbei Chen, Stephen Fairhurst, Marc Favata, Mark Hannam, Sascha Husa, Andrew Lundgren, Cole Miller, Evan Ochsner, Frank Ohme, Yi Pan, B. S. Sathyaprakash and Alan Weinstein for useful discussions and comments, and Priyanka Nayar for proofreading the manuscript. This work is supported by the LIGO Laboratory, NSF grants PHY-0653653 and PHY-0601459, NSF career grant PHY-0956189 and the David and Barbara Groce Fund at Caltech. LIGO was constructed by the California Institute of Technology and Massachusetts Institute of Technology with funding from the National Science Foundation and operates under cooperative agreement PHY-0757058. This paper has the LIGO Document Number LIGO-P1100075-v4. 

\bibliography{Spin}

\end{document}